\documentclass[aps,floatfix,showpacs,superscriptaddress,showkeys,preprint]
{revtex4}

\usepackage{titlesec}
\usepackage{graphicx,epsfig,epstopdf}
\usepackage{amssymb,amsmath,amsxtra,amsfonts}

\usepackage{bm}
\titleformat{\section}{\large\bfseries}{\thesection}{1em}{}

\newcommand{\eqa}[1]{\begin{eqnarray} #1 \end{eqnarray}}
\newcommand{\azeL}{{A_0^L}}
\newcommand{\azeR}{{A_0^R}}

\newcommand{\apeL}{{A_\perp^L}}
\newcommand{\apeR}{{A_\perp^R}}

\newcommand{\apaL}{{A_\|^L}}
\newcommand{\apaR}{{A_\|^R}}

\newcommand{\re}{{\rm Re}}
\newcommand{\im}{{\rm Im}}
\newcommand{\av}[1]{\langle #1 \rangle}
\newcommand{\bin}{{\rm bin}}

\newcommand{\bea}{\begin{eqnarray}}
\newcommand{\ena}{\end{eqnarray}}
\newcommand{\be}{\begin{equation}}
\newcommand{\en}{\end{equation}}
\newcommand{\nn}{\nonumber\\}
\newcommand{\ed}{\end{document}}

\newcommand{\Tr}{\mbox{\rm{tr}}}
\newcommand{\ord}{\mathcal{O}}

\begin{document}

\title{Decay $B\to K^\ast(\to K\pi) \ell^+ \ell^-$ in covariant quark model}

\author{S.~Dubni\v{c}ka}
\affiliation{Institute of Physics, Slovak Academy of Sciences, 
Bratislava, Slovakia} 

\author{A.Z.~Dubni\v{c}kov\'{a} }
\affiliation{Comenius University, Bratislava, Slovakia}

\author{N.~Habyl}
\email{nuigui@mail.ru}
\affiliation{Al-Farabi Kazakh National University, Almaty, Kazakhstan}

\author{M.A.~Ivanov}
\email{ivanovm@theor.jinr.ru}
\affiliation{
Joint Institute for Nuclear Research, Dubna, Russia}

\author{A.~Liptaj}
\email{andrej.liptaj@gmail.com}
\affiliation{Institute of Physics, Slovak Academy of Sciences, 
Bratislava, Slovakia} 

\author{G.S.~Nurbakova}
\email{guliya_nurbakova@mail.ru}
\affiliation{Al-Farabi Kazakh National University, Almaty, Kazakhstan}

\begin{abstract}
Our article is devoted to the study of the rare $B\to K^\ast \ell^+\ell^-$~decay
where $\ell=e,\mu,\tau$. We compute the relevant form factors in the framework 
of the covariant quark model with infrared confinement in the full kinematical 
momentum transfer region. The calculated form factors are used to evaluate
branching fractions and polarization observables in the cascade decay
$B\to K^\ast(\to K\pi)\ell^+\ell^-$. We compare the obtained results with 
available experimental data and the results from other theoretical approaches.


\end{abstract}

\maketitle

\section{Introduction}
\label{sec:intro}

The rare flavour changing neutral current (FCNC) decays 
are forbidden in the Standard Model (SM) at the tree level.
They proceed only via loops in the perturbation theory.
For this reason, these decays are sensitive to possible effects
of new physics beyond the SM. New heavy particles can contribute
to the branching fractions and the angular decay distributions.
 
It is generally believed that the decay mode $B\to K^\ast(\to K\pi)\mu^+\mu^-$
is one of the best modes to search for new physics beyond the SM. 
The angular distribution makes possible an independent measurement of several 
observables as a function of the dilepton invariant mass. 
A large number of observables obtained in this manner allows for unique 
tests of the SM contributions  (see the recent experimental review  
\cite{ChristophLangenbruchonbehalfoftheLHCb:2015iha}). A deviation of $3.3 \, \sigma$ is seen in e.g. $R(D^{\ast}) \equiv \mathcal{B}(B^- \to D^{\ast} \tau^- \bar{\nu}_{\tau}) / \mathcal{B}(B^- \to D^{\ast} \ell^- \bar{\nu}_{\ell})$ with $R(D^\ast)|_{\rm expt} = 0.321 \pm 0.021 $ \cite{Crivellin:2015hha} and $R(D^\ast)|_{\rm SM} = 0.252 \pm 0.003$ \cite{Lees:2012xj,Tanaka:2010se,Fajfer:2012vx,Kamenik:2008tj}. An analogous result questioning the lepton flavour universality exists also for $B \to K$ decays by Belle \cite{Aaij:2014ora}. 

The results of a measurement of form-factor independent angular observables 
in the decay $B^0\rightarrow K^{\ast\,0}(892)\mu^+\mu^-$ were presented 
in \cite{Aaij:2013qta}. The analysis is based on a data sample corresponding 
to an integrated luminosity of $1.0$~fb$^{-1}$, collected by 
the LHCb experiment in $p\bar{p}$ collisions at a center-of-mass energy 
of 7~TeV. Four observables are measured in six bins of the dimuon invariant 
mass squared, $q^2$, in the range $0.1 < q^2 < 19.0$~GeV$^2$. 
Agreement with the SM predictions is found for 23 of the 24 measurements. 
A local discrepancy, corresponding to $3.7$ standard
deviations, is observed in one $q^2$ bin for one of the observables. 

The measurements were followed by a large number of publications, with many 
different scenarios analyzed 
\cite{Descotes-Genon:2013wba}-\cite{Straub:2015ica}. Several authors 
discuss an appropriate choice of observables with small model dependence, 
which would discriminate between SM and new physics. Other authors use 
various assumptions or models to evaluate form factors or study 
Wilson coefficients (and related operators) in order to make theoretical 
predictions and, possibly, argue whether the observed deviations do or 
do not favor a beyond SM explanation.

Our article is devoted to the study of the $B\to K^\ast \ell^+\ell^-$ decay,
first, to evaluate the relevant form factors in the framework of the covariant
quark model with infrared confinement, and, second, to demonstrate the equivalence of our helicity-based approach with approaches of other autorhs for what concerns the model independent four-fold angular decay distribution and related results. Form factors are then exploited for evaluation 
of physical observables. The first paper in this direction has been
published in 2002, Ref.~ \cite{Faessler:2002ut}. It was one of the first paper 
where the full four-fold angular decay distribution has been derived 
for this process in terms of helicity amplitudes including lepton mass effects.
However, our model in that time was suffering from lack the of confinement
that restricted the range of applications to the hadrons (mesons and baryons) 
which satisfy the so-called "threshold inequality": the hadron mass should be smaller then the total mass of its constituents, i.e. 
the sum of the constituent quark masses. In this vein, 
our model was successfully developed for the study of light 
hadrons (e.g., pion, kaon, baryon octet, $\Delta$-resonance), 
heavy-light hadrons (e.g., $D$, $D_s$, $B$ and $B_s$-mesons, 
$\Lambda_Q$, $\Sigma_Q$, $\Xi_Q$ and $\Omega_Q$-baryons) 
and double heavy hadrons (e.g, $J/\Psi$, $\Upsilon$ and 
$B_c$-mesons, $\Xi_{QQ}$ and $\Omega_{QQ}$ baryons).
%~\cite{Ivanov:1996pz,Ivanov:1999ic}.  
To extend our approach to other hadrons we had to introduce 
extra model parameters or do some approximations, like, e.g., 
to introduce a cutoff parameter for external hadron momenta 
to guarantee the fulfilment of the above mentioned "threshold inequality". Therefore, at that stage 
we were unable to apply our approach to the study of rare decays 
involving $K^\ast$ mesons. The numerical results for the physical
observables were presented in \cite{Faessler:2002ut} only
for decays $B\to K\ell^+\ell^-$, $B_c\to D\ell^+\ell^-$
and $B_c\to D^\ast(\to D\pi)\ell^+\ell^-$.

Our relativistic constituent quark model has been refined in 2009, 
see Ref.~\cite{Branz:2009cd}, where the confinement of quarks was implemented. 
It was done, first, by introducing the scale integration in the space of 
$\alpha$-parameters, and, second, by cutting this scale integration 
on the upper limit which corresponds to an infrared cutoff. In this manner 
one removes all possible thresholds present in the initial quark diagram. 
The cutoff parameter is taken to be the same for all physical processes. 
Other model parameters were adjusted by fitting the calculated quantities 
of the basic physical processes to available experimental data. 
As an application, the electromagnetic form factors of the pion and 
the transition form factors of the omega and eta Dalitz decays have been
calculated. 

New values for the parameters of the covariant constituent quark model 
with built--in infrared confinement have been determined in 
\cite{Ivanov:2011aa} by a fit to the leptonic decay constants and 
a number of electromagnetic decays. Then the form factors of 
the $B(B_s)\to P(V)$ transitions were evaluated in the full kinematical region 
of momentum transfer squared in a parameter free way.

The model was than applied to a wide range of nonleptonic and semileptonic 
meson decays, e.g.  \cite{Dubnicka:2013vm, Issadykov:2015iba} and 
it was extended also to the baryon sector 
\cite{Gutsche:2012ze,Gutsche:2013pp,Gutsche:2013oea,Gutsche:2014zna,
Gutsche:2015mxa}. 
Different observables related to (nucleons and) heavy $\Lambda$ baryons 
were calculated in 
\cite{Gutsche:2012ze, Gutsche:2013pp, Gutsche:2013oea, Gutsche:2015mxa}. 
Furthermore, the consequences of treating the X(3872) meson as 
a tetraquark bound state were explored in 
Refs.~\cite{Dubnicka:2010kz,Dubnicka:2011mm}. 
Some of the above research was collected in the review \cite{Dineykhan:2012cp}.

Our paper is organized as follows. In Sec.~\ref{sec:model} we review salient
features of the covariant quark model.  In Sec.~\ref{sec: ff-cqm} we present
the calculated form factors in the full kinematical momentum
transfer region. In Sec.~\ref{sec:hamiltonian} we discuss the effective
Hamiltonian, matrix elements, invariant and helicity form factors.
Sec.~\ref{sec:four-fold} is devoted to the four-fold angular decay
distribution in the cascade decay $B\to K^\ast(\to K\pi)\bar\ell\ell$
and, in particular, to the relation of our helicity formalism with 
an approach based on the transversality amplitudes which is widely
used by both experimentalists and theorists. In Sec.~\ref{sec:res}
we present our numerical results on the branching fractions,
forward-backward asymmetry, longitudinal polarization and a set
of the so-called ``clean'' observables $P_i$ which depend on the hadron
uncertainties (form factors) in a minimal way. Finally, 
in Sec.~\ref{sec:summary} we summarize our results.
 
\section{Theoretical framework}
\label{sec:model}

The covariant confined quark model developed in
\cite{Efimov:1988yd,Efimov:1993ei,Faessler:2002ut,Branz:2009cd}
has been applied to a large number of elementary particle 
processes. This model can be viewed as an effective quantum field approach 
to hadronic interactions based on an interaction Lagrangian of hadrons 
interacting with their constituent quarks. The coupling strength is determined 
by the compositeness condition~$Z_H=0$ where $Z_H$ is the wave function 
renormalization constant of the hadron. The hadron field renormalization 
constant $Z_H$ characterizes the overlap between the bare hadron field and 
the bound state formed from the constituents. Once this constant is set to 
zero, the dynamics of hadron interactions is fully described by constituent 
quarks in quark loop diagrams with local constituent quark propagators.  
Matrix elements are generated by a set of quark loop diagrams 
according to the $1/N_c$ expansion. The ultraviolet divergences of the quark 
loops are regularized by including vertex functions for the hadron-quark 
vertices which, in addition, describe finite size effects due to the 
non-pointlike structure of hadrons.
Quark confinement  was implemented into the model \cite{Branz:2009cd}
by introducing an infrared cutoff on the upper limit of the scale integration 
to avoid the appearance of singularities in any matrix element. 
The infrared cutoff parameter $\lambda$ is taken to have a common 
value for all processes. The  covariant confined quark model contains 
only a few model parameters: the light and heavy constituent quark masses, 
the size  parameters that describe the size of the distribution 
of the constituent quarks inside the hadron and 
the infrared cutoff parameter $\lambda$. They are determined by a
fit  to available experimental data.

\subsection{Effective Lagrangian}
\label{subsec:efflag}

The coupling of a meson $M(q_1\bar q_2)$ to its constituent
quarks $q_1$ and $\bar q_2 $ is described  by the Lagrangian 
\be
\label{eq:Lagr}
{\mathcal L}_{\rm int}(x) = g_M M(x)\int\!\! dx_1 \!\!\int\!\!
dx_2 F_M (x,x_1,x_2)\bar q_2(x_2)\Gamma_M q_1(x_1) \, + {\rm h.c.}
\en
Here, $\Gamma_M$ is a Dirac matrix which projects onto the spin quantum 
number of the meson field $M(x)$. The function $F_M$ is related to the 
scalar part of the Bethe-Salpeter amplitude and characterizes the finite size 
of the meson. To satisfy translational invariance the function $F_H$
has to fulfill the identity $F_M(x+a,x_1+a,x_2+a)=F_M(x,x_1,x_2)$ for
any four-vector $a$. In the following we use a specific  form for the scalar 
vertex function
\be
\label{eq:vertex}
F_M(x,x_1,x_2)=\delta(x - w_1 x_1 - w_2 x_2) \Phi_M((x_1-x_2)^2),
\en
where $\Phi_M$ is the correlation function of the two constituent quarks
with masses $m_{q_1}$, $m_{q_2}$ and the mass ratios
$w_i = m_{q_i}/(m_{q_1}+m_{q_2})$.

We choose a simple Gaussian form of the vertex function $\bar \Phi_M(-\,k^2)$. 
The minus sign in the argument of this function is chosen to emphasize 
that we are working in the Minkowski space. One has
\be
\bar \Phi_M(-\,k^2) 
= \exp\left(k^2/\Lambda_M^2\right),
\label{eq:Gauss}
\en
where the parameter $\Lambda_M$ characterizes the size of the meson.
Since $k^2$ turns into $-\,k_E^2$ in the Euclidean space, the form
(\ref{eq:Gauss}) has the appropriate fall-off behavior in the Euclidean region.
We emphasize that any choice for  $\Phi_M$ is appropriate
as long as it falls off sufficiently fast in the ultraviolet region of
the Euclidean space to render the corresponding Feynman diagrams ultraviolet 
finite. We choose a Gaussian form for $\Phi_M$ for calculational convenience.

\subsection{Compositeness condition}
\label{subsec:Z=0}

The coupling constant $g_M$ in Eq.~(\ref{eq:Lagr}) is determined by the
so-called {\it compositeness condition} suggested by  
Weinberg~\cite{Weinberg:1962hj} and Salam~\cite{Salam:1962ap} 
(for a review, see Ref.~\cite{Hayashi:1967hk}) and extensively used in 
our studies (for details, see Ref.~\cite{Efimov:1993ei}). 
The compositeness condition requires that the renormalization constant $Z_M$ 
of the elementary meson field $M(x)$ is set to zero, i.e.,
\be
\label{eq:Z=0}
Z_M \, = \, 1 - \, \frac{3g^2_M}{4\pi^2} \,\bar\Pi'_M(m^2_M) \, = \, 0,
\en
where $\bar\Pi^\prime_H$ is the derivative of the meson mass operator.
To clarify the physical meaning of the compositeness  condition 
in Eq.~(\ref{eq:Z=0}), we first want to remind the reader
that the renormalization constant $Z_M^{1/2}$ can be also interpreted as
the matrix element between the physical and the corresponding bare state.
The condition  $Z_M=0$ implies that the physical state does not contain
the bare state and is appropriately described as a bound state.
The interaction Lagrangian of Eq.~(\ref{eq:Lagr}) and
the corresponding free parts of the Lagrangian describe
both the constituents (quarks) and the physical particles (hadrons)
which are viewed as the bound states of the quarks.
As a result of the interaction, the physical particle is dressed,
i.e. its mass and wave function have to be renormalized.
The condition $Z_M=0$ also effectively excludes
the constituent degrees of freedom from the space of physical states.
It thereby guarantees that there is no double counting
for the physical observable under consideration.
The constituents exist only in  virtual states.
One of the corollaries of the compositeness condition is the absence
of a direct interaction of the dressed charged particle with the
electromagnetic field. Taking into account both the tree-level
diagram and the diagrams with the self-energy insertions into the
external legs (i.e. the tree-level diagram times $Z_M -1$) yields
a common factor $Z_M$  which is equal to zero. 

\begin{figure*}[htbp]
\begin{center}
\epsfig{figure=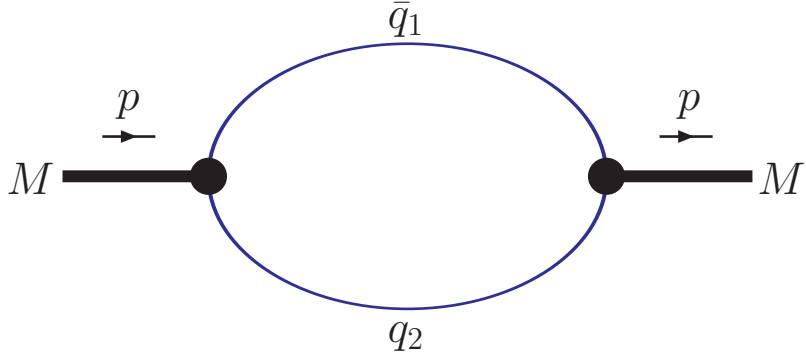,scale=1.}
\caption{Diagram describing the meson mass function.}
\label{fig:mass}
\end{center}
\end{figure*}

The diagram describing the meson mass function is shown in 
Fig.~(\ref{fig:mass}). The derivative of the mass function is calculated using 
the identity

\be
\frac{d}{dp^2} =\frac{1}{2p^2} p^\mu \frac{d}{dp^\mu}.
\en
In the case of the pseudoscalar and vector mesons
the derivatives of the meson mass operator appearing in Eq.~(\ref{eq:Z=0})
are written as

\bea
\widetilde\Pi'_P(p^2) &=&
\frac{1}{2p^2}\,
p^\alpha\frac{d}{dp^\alpha}\,
\int\!\! \frac{d^4k}{4\pi^2i}\, \widetilde\Phi^2_P(-k^2)\,
{\rm tr} \biggl[\gamma^5 S_1(k+w_1 p)\gamma^5 S_2(k-w_2 p) \biggr] 
\label{eq:Mass-operator-P}\\
&=&
\frac{1}{2p^2}\,\int\!\! \frac{d^4k}{4\pi^2i}\,\widetilde\Phi^2_P(-k^2)\,
\Big\{
w_1\,{\rm tr} 
\biggl[\gamma^5 S_1(k+w_1 p)\!\not\!p\, S_1(k+w_1 p)\gamma^5 S_2(k-w_2 p) \biggr] 
\nn
&&
\phantom{
\frac{1}{2p^2}\,\int\!\! \frac{d^4k}{4\pi^2i}\,\widetilde\Phi^2_P(-k^2)\,
\!\!\!
}
-w_2\,{\rm tr} 
\biggl[\gamma^5 S_1(k+w_1 p)\gamma^5 S_2(k-w_2 p)\!\not\!p\, S_2(k-w_2 p) \biggr]
\Big\} ,
\nn[2ex]
 \tilde\Pi'_V(p^2)  &=&
\frac{1}{3}\left( g_{\mu\nu}-\frac{p_\mu p_\nu}{p^2}\right)
\nn
&\times&
\frac{1}{2p^2}\,p^\alpha\frac{d}{dp^\alpha}\,
\int\!\! \frac{d^4k}{4\pi^2i}\, \widetilde\Phi^2_V(-k^2)\,
{\rm tr} \biggl[\gamma^\mu S_1(k+w_1 p) \gamma^\nu S_2(k-w_2 p) \biggr]
\label{eq:Mass-operator-V}\\
&=&
\frac{1}{3}\left( g_{\mu\nu}-\frac{p_\mu p_\nu}{p^2}\right)
\nn
&\times&
\frac{1}{2p^2}\,\int\!\! \frac{d^4k}{4\pi^2i}\,\widetilde\Phi^2_V(-k^2)\,
\Big\{
w_1\,{\rm tr} 
\biggl[
\gamma^\mu S_1(k+w_1 p)\!\not\!p\, S_1(k+w_1 p)\gamma^\nu S_2(k-w_2 p) 
\biggr] 
\nn
&&
\phantom{
\frac{1}{2p^2}\,\int\!\! \frac{d^4k}{4\pi^2i}\,\widetilde\Phi^2_V(-k^2)\,
\!\!\!
}
-w_2\,{\rm tr} 
\biggl[
\gamma^\mu S_1(k+w_1 p)\gamma^\nu S_2(k-w_2 p)\!\not\!p\, S_2(k-w_2 p) \biggr]
\Big\},
\nonumber 
\ena
where $\widetilde\Phi_H(-k^2)$ is the Fourier-transform of
the vertex function defined by Eq.~(\ref{eq:Gauss}).
We use free fermion propagators for the quarks given by
\be
\label{eq:quark-prop}
S_i(k)=\frac{1}{m_{q_i}-\not\! k}
\en
with an effective constituent quark mass $m_{q_i}$. 

\subsection{Infrared confinement}
\label{Sec:IR} 

In the  paper \cite{Branz:2009cd} we have included the confinement of quarks 
to our model.
It was done, first, by introducing the scale integration
in the space of $\alpha$-parameters, and, second, by cutting this 
scale integration on the upper limit which corresponds to an infrared cutoff. 
In this manner one removes all possible thresholds 
present in the initial quark diagram. 
The cutoff parameter is taken to be the same for all physical processes.
We have adjusted other model parameters by fitting the calculated
quantities of the basic physical processes to available experimental data.
In the papers \cite{Dubnicka:2010kz,Dubnicka:2011mm}
we have applied the developed approach to the 4-body system - tetraquark 
X(3872). 

Let us give the basic features of the infrared confinement in our model.
All physical matrix elements are described by the Feynman diagrams which
are the convolution of the free quark propagators and vertex functions. 
 Let $n$, $\ell$ and  $m$ be the number
of the propagators, loops and vertices, respectively.
In the Minkowski space the $\ell$-loop diagram will be represented as
\bea
&&
\Pi(p_1,...,p_m) = 
\int\!\! [d^4k]^\ell  
\prod\limits_{i_1=1}^{m} \,
\Phi_{i_1+n} \left( -K^2_{i_1+n}\right)
\prod\limits_{i_3=1}^n\, S_{i_3}(\tilde k_{i_3}+v_{i_3}),
\nn
&&\nn
&&
K^2_{i_1+n} =\sum_{i_2}(\tilde k^{(i_2)}_{i_1+n}+v^{(i_2)}_{i_1+n})^2,
\label{eq:diag}
\ena
where the vectors $\tilde k_i$  are  linear combinations 
of the loop momenta $k_i$. The $v_i$ are  linear combinations 
of the external momenta $p_i$ to be specified in what follows.
The strings of Dirac matrices appearing in the calculation need not concern 
us since they do not depend on the momenta. 
The external momenta $p_i$ are all chosen to be ingoing such that one has 
$\sum\limits_{i=1}^m p_i=0$. All calculations proceed in the Euclidean
region both for the loop momenta $k_i$ and the external momenta $p_i$
so that $k^2_i\le 0$,  $p^2_i\le 0$.  
 
Using the Schwinger representation of the local quark propagator one has
\be
S(k) = (m+\not\! k)
\int\limits_0^\infty\! 
d\alpha\,e^{-\alpha\,(m^2-k^2)}\,.
\en
For the vertex functions one takes the Gaussian form
\be
\label{eq:vert} 
\Phi_{i+n} \left( -K^2\right)\,
 =
\exp\left[\alpha_{i+n}\,K^2\right] \qquad i=1,...,m\, ,
\en
where the parameters $\alpha_{i+n}=s_{i}=1/\Lambda^2_{i}$ are 
related to the size parameters. The integrand in Eq.~(\ref{eq:diag})
has a Gaussian form with the exponential $kak+2kr+R$ where $a$ is
$\ell\times\ell$ matrix depending on the parameter $\alpha_i$,
$r$ is the $\ell$-vector composed from the external momenta, and
$R$ is a quadratic form of the external momenta.
Tensor loop integrals are calculated with the help of the differential
representation 
\be
k_i^\mu e^{2kr} = \frac{1}{2}\frac{\partial}{\partial r_{i\,\mu}}e^{2kr}.
\en 
This allows to use the  operator identity
\be
\int\! d^{4}k\,P\left(k\right)e^{2kr}
=\int\! d^{4}k\, P\left(\frac{1}{2}\frac{\partial}{\partial r}\right)e^{2kr}
=P\left(\frac{1}{2}\frac{\partial}{\partial r}\right)\int d^{4}k\: e^{2kr}
\en
which is written for one loop integration.
The second identity then reads
\be
\int\limits_0^\infty d^{n}\alpha\: 
P\left(\frac12\frac{\partial}{\partial r}\right) e^{-r^2/a}
=
\int\limits_0^\infty d^{n}\alpha\:
 e^{-r^2/a}
P\left(\frac12\frac{\partial}{\partial r}-\frac{r}{a}\right),
\en
where $\mathrm{r=r\left(\alpha_{i}\right)}$ and 
$a=a\mathrm{\left(\Lambda_{M},\alpha_{i}\right)}$. It
simplifies the computation following the trace evaluation: the polynomial
in the derivative operator which results from the trace can be applied
to an identity, instead being applied to a more complicated exponential
function.

We have written a FORM \cite{Vermaseren:2000nd} program that performs 
the necessary commutations of the differential operators in a very efficient 
way. After doing the loop integrations one obtains
\be
\Pi =  \int\limits_0^\infty d^n \alpha \, F(\alpha_1,\ldots,\alpha_n) \,,
\en
where $F$ stands for the whole structure of a given diagram. 
The region over which the set of Schwinger parameters $\alpha_i$ is integrated can be turned into a simplex by 
introducing an additional $t$--integration via the identity 
\be 
1 = \int\limits_0^\infty dt \, \delta(t - \sum\limits_{i=1}^n \alpha_i)
\en 
leading to 
\be
\hspace*{-0.2cm}
\Pi   = \int\limits_0^\infty\! dt t^{n-1}\!\! \int\limits_0^1\! d^n \alpha \, 
\delta\Big(1 - \sum\limits_{i=1}^n \alpha_i \Big) \, 
F(t\alpha_1,\ldots,t\alpha_n). 
\label{eq:loop_2} 
\en
There are altogether $n$ numerical integrations: $(n-1)$ $\alpha$--parameter
integrations and the integration over the scale parameter $t$. 
The very large $t$-region corresponds to the region where the singularities
of the diagram with its local quark propagators start appearing. 
However, as described in \cite{Branz:2009cd}, if one introduces 
an infrared cutoff on the upper limit of the $t$-integration, all 
singularities vanish because the integral is now convergent for any value
of the set of kinematic variables.
We cut off the upper integration at $1/\lambda^2$ and obtain
\be
\hspace*{-0.2cm}
  \Pi^c = \!\!  
\int\limits_0^{1/\lambda^2}\!\! dt t^{n-1}\!\! \int\limits_0^1\! d^n \alpha \, 
\delta\Big(1 - \sum\limits_{i=1}^n \alpha_i \Big) \, 
F(t\alpha_1,\ldots,t\alpha_n).
\en  
By introducing the infrared cutoff one has removed all potential thresholds 
in the quark loop diagram, i.e. the quarks are never on-shell and are thus
effectively confined. 
%%% insertion
Similar ideas have also been pursued in Refs.~\cite{Ebert:1996vx} where 
an infrared cutoff had been introduced in the context of 
a Nambu-Jona-Lasinio model by employing a proper time regularization.
Such approach found some applications, in particular,  for spectroscopy of 
charmonia \cite{Bedolla:2015mpa}. Since the contact interaction has been
employed one has to introduce also the ultraviolet cut-off. 
The values of both  infrared and  ultraviolet cut-off parameters
are supposed to be different for different hadrons. 

We take the infrared cutoff parameter $\lambda$ to be the 
same in all physical processes. It may be understood from a proper time
regularization of the generating functional describing the NJL model
where a proper time is general for all matrix elements. Our 
``$t$''-integration parameter is analogous of a proper time and, hence, 
should not depend on the hadron characteristics of the certain process. 
It is important that we use nonlocal interactions of hadrons with their 
constituents.
The relevant vertex functions characterize the properties of the hadron  
in such a way that the size parameters appearing in Eq.~(\ref{eq:Gauss})
are different for different hadrons. Their values are determined
by fitting the calculated observables to the experimental data.

%%% end insertion

\section{ \boldmath{$B-K^\ast$} form factors in the covariant quark
model}
\label{sec: ff-cqm}

\begin{figure*}[htbp]
\begin{center}
\includegraphics[width=0.90\textwidth]{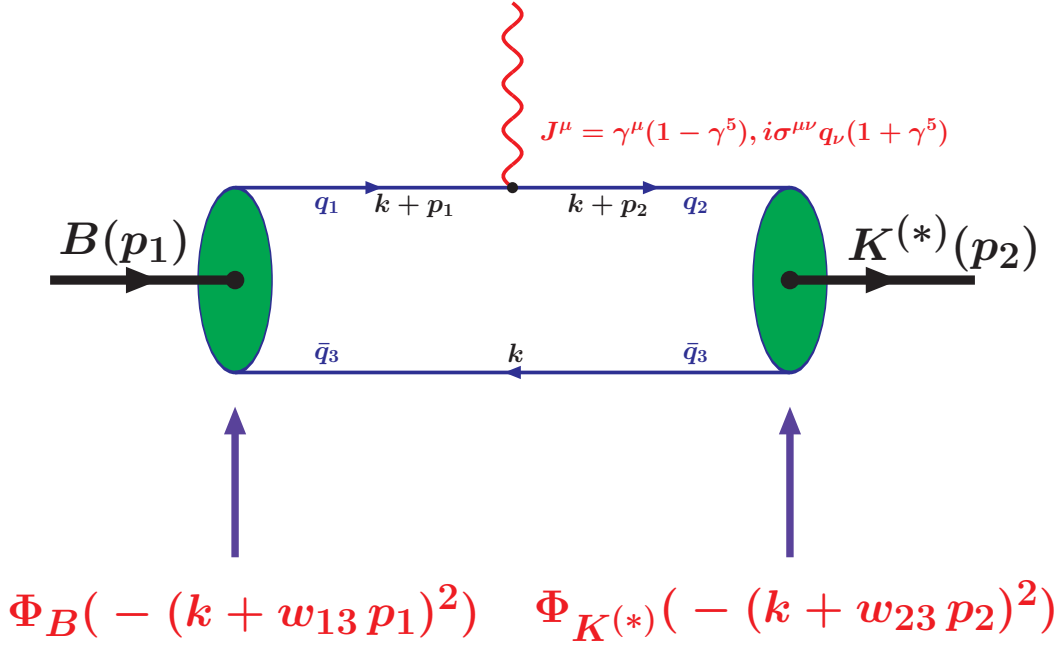} 
\caption{Diagrammatic representation of the matrix elements
describing $B\to K^\ast$ transitions.}
\label{fig:diag}
\end{center}
\end{figure*}

Herein our primary subject is the matrix element
which is described by the Feynman diagram shown in Fig.~\ref{fig:diag}.
It can be expressed via dimensionless form factors:
\bea
&&
\langle 
V(p_2,\epsilon_2)_{[\bar q_1 q_3]}\,
|\,\bar q_2\, O^{\,\mu}\,q_1\, |\,P_{[\bar q_3 q_2]}(p_1)
\rangle 
\,=\,
\nn
&=&
N_c\, g_P\,g_V \!\! \int\!\! \frac{d^4k}{ (2\pi)^4 i}\, 
\widetilde\Phi_P\Big(-(k+w_{13} p_1)^2\Big)\,
\widetilde\Phi_V\Big(-(k+w_{23} p_2)^2\Big)
\nn
&\times&
{\rm tr} \biggl[ 
O^{\,\mu} \,S_1(k+p_1)\,\gamma^5\, S_3(k) \not\!\epsilon_2^{\,\,\dagger} \,
S_2(k+p_2)\, \biggr]
\nn
 & = &
\frac{\epsilon^{\,\dagger}_{\,\nu}}{m_1+m_2}\,
\Big( - g^{\mu\nu}\,P\cdot q\,A_0(q^2) + P^{\,\mu}\,P^{\,\nu}\,A_+(q^2)
       + q^{\,\mu}\,P^{\,\nu}\,A_-(q^2)
\nn 
&& + i\,\varepsilon^{\mu\nu\alpha\beta}\,P_\alpha\,q_\beta\,V(q^2)\Big),
\label{eq:PV}\\[3ex]
&&
\langle 
V(p_2,\epsilon_2)_{[\bar q_1 q_3]}\,
|\,\bar q_2\, (\sigma^{\,\mu\nu}q_\nu(1+\gamma^5))\,q_1\, |\,P_{[\bar q_3 q_2]}(p_1)
\rangle 
\,=\,
\nn
&=&
N_c\, g_P\,g_V \!\! \int\!\! \frac{d^4k}{ (2\pi)^4 i}\, 
\widetilde\Phi_P\Big(-(k+w_{13} p_1)^2\Big)\,
\widetilde\Phi_V\Big(-(k+w_{23} p_2)^2\Big)
\nn
&\times&
{\rm tr} \biggl[ 
(\sigma^{\,\mu\nu}q_\nu(1+\gamma^5))
\,S_1(k+p_1)\,\gamma^5\, S_3(k) \not\!\epsilon_2^{\,\,\dagger} \,S_2(k+p_2)\, 
\biggr]
\nn
 & = &
\epsilon^{\,\dagger}_{\,\nu}\,
\Big( - (g^{\mu\nu}-q^{\,\mu}q^{\,\nu}/q^2)\,P\cdot q\,a_0(q^2) 
       + (P^{\,\mu}\,P^{\,\nu}-q^{\,\mu}\,P^{\,\nu}\,P\cdot q/q^2)\,a_+(q^2)
\nn
&&
+ i\,\varepsilon^{\mu\nu\alpha\beta}\,P_\alpha\,q_\beta\,g(q^2)\Big).
\label{eq:PVT}
\ena

Here, $P=p_1+p_2$, $q=p_1-p_2$, $\epsilon_2^\dagger\cdot p_2=0$,
$p_i^2=m_i^2$. Since there are three sorts of quarks involved in these
processes, we introduce the notation with two subscripts
$w_{ij}=m_{q_j}/(m_{q_i}+m_{q_j})$ $(i,j=1,2,3)$ so that $w_{ij}+w_{ji}=1$. 
The form factors defined in Eq.\,(\ref{eq:PVT}) satisfy the physical 
requirement $a_0(0)=a_+(0)$, which ensures that no kinematic singularity 
appears in the matrix element at $q^2=0$ GeV.

We will use the latest fit done in Ref.~\cite{Ganbold:2014pua}.
The  fitted values of the constituent quark masses $m_q$, the infrared cutoff
$\lambda$, and the size parameters $\Lambda_H$ are given by
Eq.~(\ref{eq:fitmas}) and Table~\ref{tab:Lambda-H}. 

\be
\def\arraystretch{1.5}
\begin{array}{cccccc}
     m_{u/d}        &      m_s        &      m_c       &     m_b & \lambda  &   
\\\hline
 \ \ 0.241\ \   &  \ \ 0.428\ \   &  \ \ 1.67\ \   &  \ \ 5.05\ \   & 
\ \ 0.181\ \   & \ {\rm GeV} 
\end{array}
\label{eq:fitmas}
\en

\begin{table}[ht]
\caption{The fitted values of the size parameters  $\Lambda_M$ in GeV.}
\label{tab:Lambda-H}
\begin{center}
\def\arraystretch{1.2}
\begin{tabular}{cccccccccc}
\hline
 $\pi$ & $K$  & $D$  & $D_s$ & $B$  & $B_s$ & $B_c$ & $\eta_c$ & $\eta_b$ &\\ 
%\hline
  0.87 & 1.02 & 1.71 & 1.81  & 1.96 & 2.05  & 2.50 & 2.06 & 2.95 &\\
\hline\hline
 $\rho$ & $\omega$ & $\phi$ & $J/\psi$ & $K^\ast$ & $D^\ast$ & $D_s^\ast$ & 
  $B^\ast$ & $B_s^\ast$ & $\Upsilon$ \\ 
%\hline
 \ \ 0.61\ \  & \ \  0.50 \ \  & \ \  0.91\ \  & \ \  1.93\ \  & \ \  0.75\ \  
 & \ \  1.51\ \  & \ \  1.71 \ \  & \ \  1.76\ \  & \ \  1.71 \ \ & 
\ \ 2.96 \ \ \\ 
\hline
\end{tabular}
\end{center}
\end{table}

\begin{figure*}[htbp]
\begin{center}
%\vspace*{.25cm}
\epsfig{figure=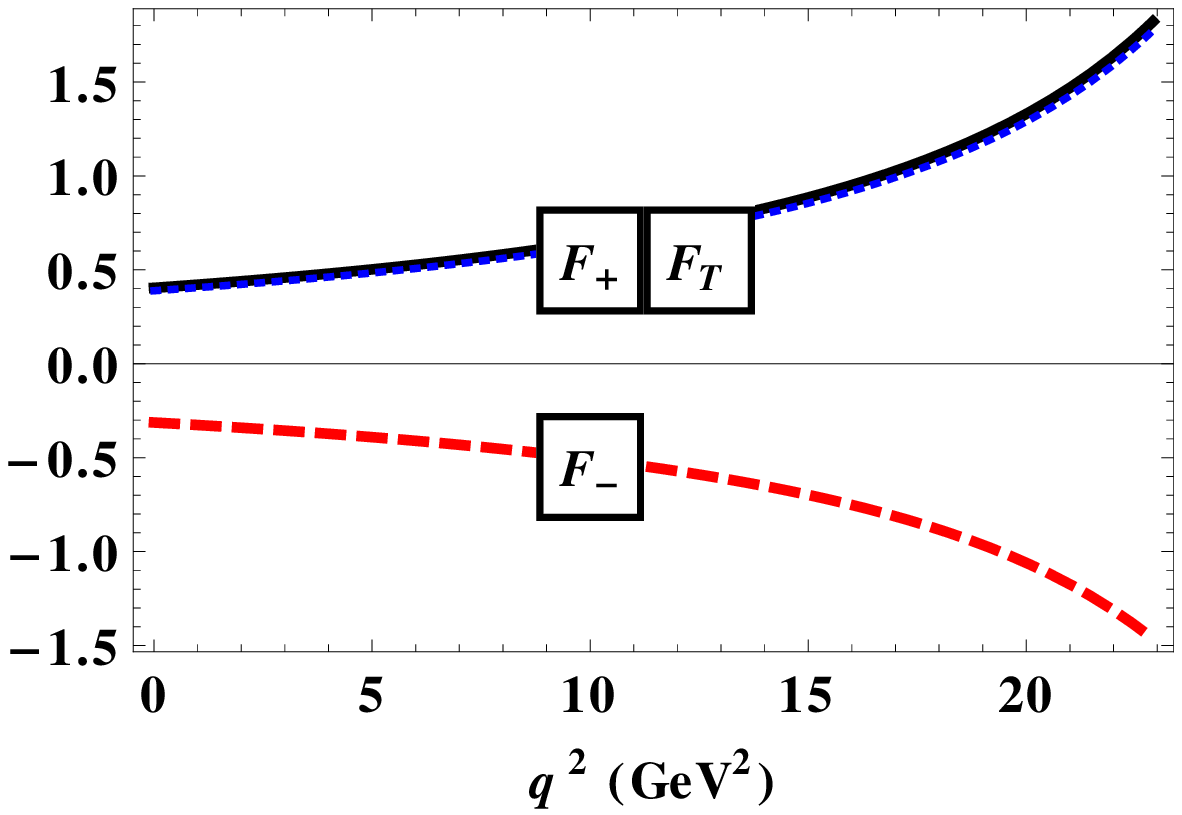,scale=1.}
\caption{The  $q^{2}$-dependence of the $B-K$~transition form factors.
\label{fig:BK}
}
\end{center}
\end{figure*}

\begin{figure*}[htbp]
\begin{center}
%\vspace*{.25cm}
\epsfig{figure=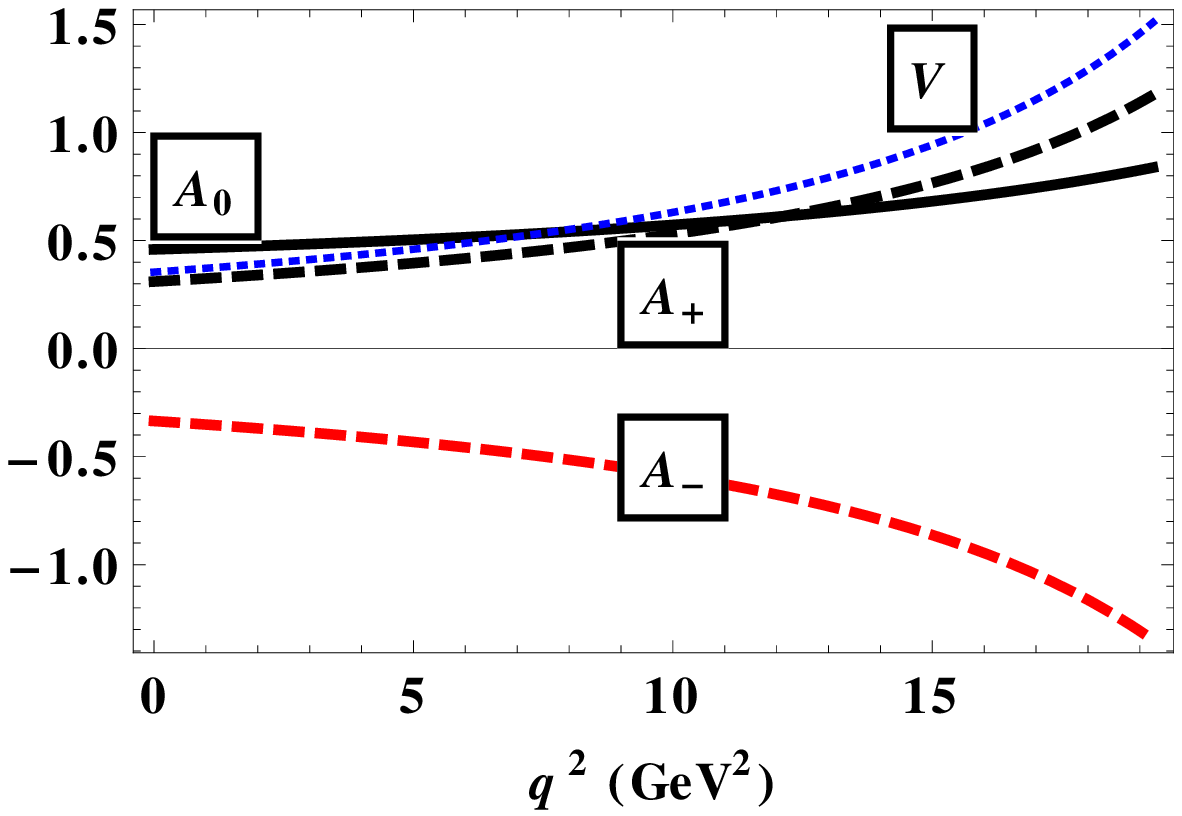,scale=1}
\caption{The  $q^{2}$-dependence of the 
$B-K^\ast$~transition $V,A$ form factors.
\label{fig:BKv1}
}
\end{center}
\end{figure*}

\begin{figure*}[htbp]
\begin{center}
%\vspace*{.25cm}
\epsfig{figure=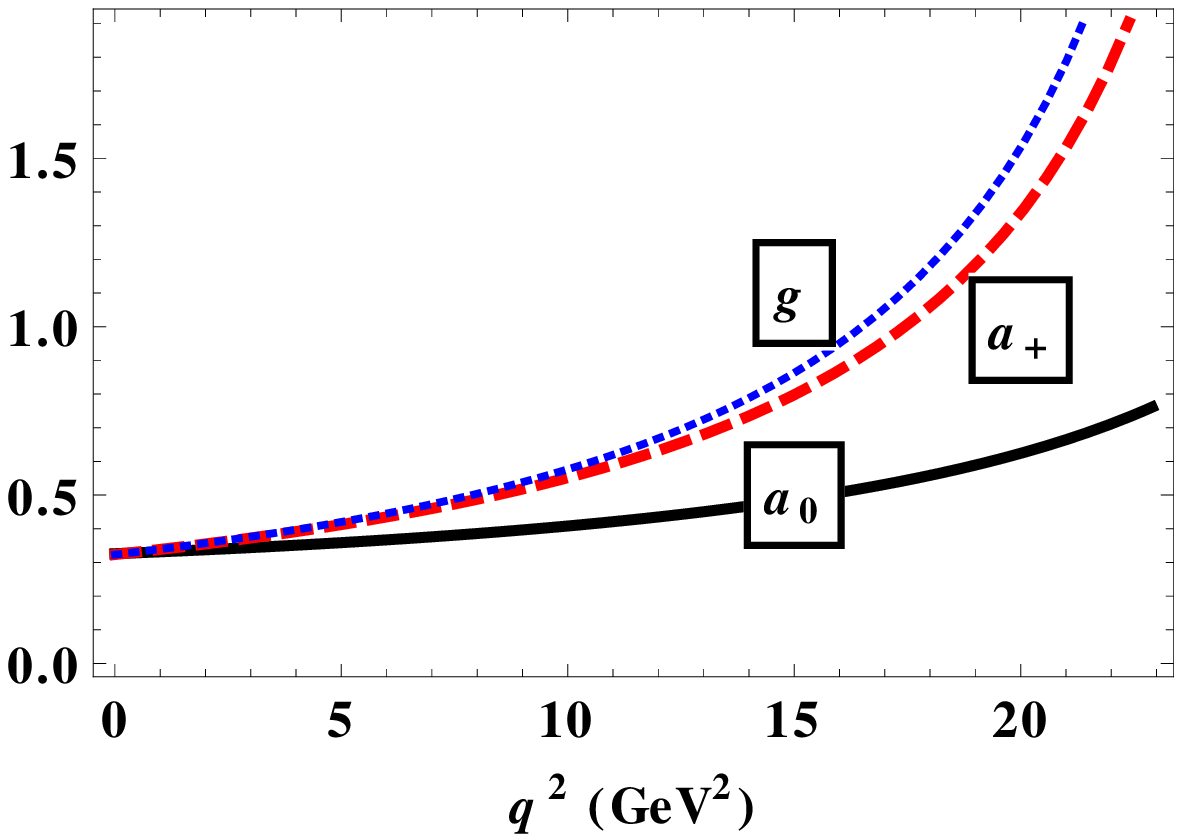,scale=1}
\caption{The  $q^{2}$-dependence of the 
$B-K^\ast$~transition $T$ form factors.
\label{fig:BKv2}
}
\end{center}
\end{figure*}

Our form factors are represented as three-fold integrals which
are calculated by using NAG routines. They are shown in Figs.~\ref{fig:BK},
\ref{fig:BKv1} and \ref{fig:BKv2}. 
The results of our numerical calculations are well approximated
by the parametrization
\be
F(q^2)=\frac{F(0)}{1-a s+b s^2}\,, \qquad s=\frac{q^2}{m_1^2}\,.
\label{eq:ff_approx}
\en
The values of $F(0)$, $a$, and $b$ are listed  in Table~\ref{tab:BKvff}.
%%% New Insertion 1
One can compare the obtained values of form factors at maximum recoil
$q^2=0$ with those obtained in other approaches, see, for example,
Table IV in \cite{Ivanov:2007cw} and references therein.
%%% End New Insertion 1
%
\begin{table}[ht]
\caption{Parameters for the approximated form factors
in Eqs.~(\ref{eq:ff_approx}) of  the $B \to K^\ast$ transitions.} 
\begin{center}
\begin{tabular}{ccccc|ccc}
\hline
&\qquad $A_0$ \qquad &\qquad $A_+$ \qquad  &\qquad $A_-$ \qquad &\qquad $V$ 
\qquad 
&\qquad $a_0$ \qquad &\qquad $a_+$ \qquad &\qquad $g$ \qquad \\
\hline
$F(0)$ &  0.459   & 0.310 & $-0.335$ & 0.354 & 0.326    & 0.323 & 0.323 \\
$a$    &  0.439   & 1.252 & 1.306    & 1.345 & 0.457    & 1.249 & 1.350 \\
$b$    & $-0.311$ & 0.270 & 0.316    & 0.343 & $-0.290$ & 0.268 & 0.349 \\
\hline
\end{tabular}
\label{tab:BKvff} 
\end{center}
\end{table}

\clearpage
\section{Effective Hamiltonian, invariant amplitudes and decay distributions}
\label{sec:hamiltonian}

The rare exclusive decays are described by the effective Hamiltonian 
obtained from the SM-diagrams using the operator product expansion and 
renormalization group techniques. It allows one to separate the short-distance 
contributions and isolate  them in the Wilson coefficients which can be 
studied systematically within perturbative QCD. The long-distance
contributions are contained in the matrix elements of local operators.
Contrary to the short-distance contributions the calculation of
such matrix elements requires  nonperturbative methods and is therefore
model dependent.

The rare decay $b \to s(d) \ell^+ \ell^-$ can be described in terms of 
the effective Hamiltonian \cite{Buchalla:1995vs}:  
\be
{\mathcal H}_{\rm eff} = - \frac{4G_F}{\sqrt{2}} \lambda_t   
              \sum_{i=1}^{10} C_i(\mu)  \ord_i(\mu) ,
\label{eq:effHam}
\en
where $C_i(\mu)$ and $\ord_i(\mu)$ are the Wilson coefficients and local 
operators, respectively. $\lambda_t \equiv |V_{ts}^\ast V_{tb}|$ is the product 
of CKM matrix elements. The standard set \cite{Buchalla:1995vs} 
of local operators for $b \to s l^+ l^-$ transition is written as 
\bea 
\begin{array}{ll} 
\ord_1     =  (\bar{s}_{a_1}\gamma^\mu P_L c_{a_2})
              (\bar{c}_{a_2}\gamma_\mu P_L b_{a_1}),                   &
\ord_2     =  (\bar{s}\gamma^\mu P_L c)  (\bar{c}\gamma_\mu P_L b),   
\\[2ex]
\ord_3     =  (\bar{s}\gamma^\mu P_L  b) \sum_q(\bar{q}\gamma_\mu P_L q),  &
\ord_4     =  (\bar{s}_{a_1}\gamma^\mu P_L  b_{a_2}) 
              \sum_q (\bar{q}_{a_2}\gamma_\mu P_L q_{a_1}),
\\[2ex]
\ord_5     =  (\bar{s}\gamma^\mu P_R b)_{V-A}
              \sum_q(\bar{q}\gamma_\mu P_R q),            &
\ord_6     =  (\bar{s}_{a_1}\gamma^\mu P_R b_{a_2 }) 
              \sum_q  (\bar{q}_{a_2} \gamma_\mu P_R q_{a_1}),               
\\[2ex]
\ord_7     =  \frac{e}{16\pi^2} \bar m_b\, 
              (\bar{s} \sigma^{\mu\nu} P_R b) F_{\mu\nu},       &
\ord_8    =  \frac{g}{16\pi^2} \bar m_b\, 
              (\bar{s}_{a_1} \sigma^{\mu\nu} P_R {\bf T}_{a_1a_2} b_{a_2}) 
              {\bf G}_{\mu\nu},            
\\[2ex]
\ord_9     = \frac{e^2}{16\pi^2}        
             (\bar{s} \gamma^\mu P_L b) (\bar\ell\gamma_\mu \ell),     &
\ord_{10}  = \frac{e^2}{16\pi^2} 
             (\bar{s} \gamma^\mu P_L b)  (\bar\ell\gamma_\mu\gamma_5 \ell), 
\end{array}
\label{eq:operators}
\ena
where ${\bf G}_{\mu\nu}$ and $F_{\mu\nu}$ are the gluon and photon 
field strengths, respectively; ${\bf T}_{a_1a_2}$ are the generators of 
the $SU(3)$ color group; $a_1$ and $a_2$ denote color indices 
(they are omitted in the color-singlet currents).
The chirality projection operators are 
$P_{L,R} = (1 \mp \gamma_5)/2$ and $\mu$ is a renormalization scale.
$\ord_{1,2}$ are current-current operators, 
$\ord_{3-6}$ are QCD penguin operators,  $\ord_{7,8}$ are "magnetic  
penguin" operators, and $\ord_{9,10}$ are semileptonic electroweak penguin 
operators.

By using the effective Hamiltonian defined by Eq.~(\ref{eq:effHam})
one can write the matrix element of the exclusive transition 
$B\to K^\ast \ell^+ \ell^-$ as  
\bea
{\mathcal M} & = & 
\frac{G_F}{\sqrt{2}}\cdot\frac{ \alpha\lambda_t}{\pi} \cdot
\Big\{
C_9^{\rm eff}\,<K^\ast\,|\,\bar{s}\,\gamma^\mu\, P_L\, b\,|\,B> 
\left( \bar \ell \gamma_\mu \ell \right)
\nn
&-& \frac{2\bar m_b}{q^2}\,C_7^{\rm eff}\, 
<K^\ast\,|\,  \bar{s}\,i\sigma^{\mu \nu} q_\nu \,P_R\, b\, |\,B>  
\left( \bar \ell \gamma_\mu \ell \right)
\nn
&+& C_{10}\, <K^\ast\,|\,\bar{s}\,\gamma^\mu P_L \, b\,|\,B>
\left(\bar \ell \gamma_\mu \gamma_5 \ell\right)
\Big\},
\label{eq:matrix-elem}
\ena
where $C_7^{\rm eff}= C_7 -C_5/3 -C_6$.
% 
%---------- insertion -----------------------
%
One has to note that matrix element in Eq.(\ref{eq:matrix-elem}) contains
both a free quark decay amplitude coming from the operators
$\ord_7$, $\ord_9$ and $\ord_{10}$ (gluon magnetic penquin  $\ord_{8}$
does not contribute) and, in addition, certain long-distance effects 
from the matrix elements of four-quark operators $\ord_i\,\,(i=1,\ldots,6)$
which usually are absorbed into a redefinition of the short-distance 
Wilson-coefficients.
%
%-------- end insertion -------------
%
The Wilson coefficient 
$ C_9^{\rm eff}$ effectively takes into account, first, the contributions 
from the four-quark operators $\ord_i$ ($i=1,...,6$) and, second, 
the nonperturbative effects coming from the $c\bar c$-resonance contributions
which are as usual parametrized by the Breit-Wigner ansatz \cite{Ali:1991is}:
\bea
C_9^{\rm eff} & = & C_9 + 
C_0 \left\{
h(\hat m_c,  s)+ \frac{3 \pi}{\alpha^2}\,  \kappa\,
         \sum\limits_{V_i = \psi(1s),\psi(2s)}
      \frac{\Gamma(V_i \rightarrow l^+ l^-)\, m_{V_i}}
{  {m_{V_i}}^2 - q^2  - i m_{V_i} \Gamma_{V_i}}
\right\} 
\nn
&-& \frac{1}{2} h(1,  s) \left( 4 C_3 + 4 C_4 +3 C_5 + C_6\right)  
\nn
&-& \frac{1}{2} h(0,  s) \left( C_3 + 3 C_4 \right) +
\frac{2}{9} \left( 3 C_3 + C_4 + 3 C_5 + C_6 \right)\,,
\label{eq:C9eff}
\ena
where $C_0\equiv 3 C_1 + C_2 + 3 C_3 + C_4+ 3 C_5 + C_6$.
Here 
%%% New insertion 2
the charm-loop function is written as
%%% end  New insertion 2
\bea 
h(\hat m_c,  s) & = & - \frac{8}{9}\ln\frac{\bar m_b}{\mu} 
- \frac{8}{9}\ln\hat m_c +
\frac{8}{27} + \frac{4}{9} x 
\nn
& - & \frac{2}{9} (2+x) |1-x|^{1/2} \left\{
\begin{array}{ll}
\left( \ln\left| \frac{\sqrt{1-x} + 1}{\sqrt{1-x} - 1}\right| - i\pi 
\right), &
\mbox{for } x \equiv \frac{4 \hat m_c^2}{ s} < 1, \nonumber \\
 & \\
2 \arctan \frac{1}{\sqrt{x-1}}, & \mbox{for } x \equiv \frac
{4 \hat m_c^2}{ s} > 1,
\end{array}
\right. 
\nn
h(0,  s) & = & \frac{8}{27} -\frac{8}{9} \ln\frac{\bar m_b}{\mu} - 
\frac{4}{9} \ln s + \frac{4}{9} i\pi,
\nonumber 
\ena
where $\hat m_c=\bar m_c/m_B$, $s=q^2/m_B^2$ and $\kappa=1/C_0$.
In what follows we drop the charm resonance contributions
by putting  $\kappa=0$. We denote the QCD quark masses by the bar symbol
to distinguish them from the constituent quark masses used in the model,
see Eq.~(\ref{eq:quark-prop}). We will use the value of $\mu=\bar m_{b\,\rm pole}$
for the renormalization scale. The numerical values for the model-independent
input parameters and the corresponding values of the Wilson coefficients
are given in Table~\ref{tab:input}.
%%% New insertion 3
Besides the charm-loop perturbative contribution, two loop contributions have
been calculated in \cite{Greub:2008cy}. A global analysis
of $b\to s\ell\ell$ anomalies has been performed in 
Ref.~\cite{Descotes-Genon:2015uva} with the NNLL corrections included.
It was shown that they amount up to 15$\%$. The discussion of 
the non-local $c\bar c$ contributions maybe also found in
Ref.~\cite{Beylich:2011aq}.
%%% End New insertion 3

We specify our choice of the momenta for the reaction
$B(p_1)\to H_{\rm out}(p_2)+\ell^+(k_1) + \ell^-(k_2)$
as $p_1=p_2+k_1+k_2$ with $p_1^2=m_1^2$, $p_2^2=m_2^2$ and  
$k_1^2=k_2^2=m_\ell^2$ where $k_1$ and 
$k_2$ are the $l^+$ and $l^-$ momenta, and $m_1$, $m_2$, $m_\ell$ are 
the masses of the initial $B$ meson, final meson $H_{\rm out}=K,K^\ast$ 
and lepton, respectively.

The matrix element in Eq~(\ref{eq:matrix-elem}) can now be written 

\be
{\mathcal M}=
\frac{G_F}{\sqrt{2}}\cdot\frac{\alpha\lambda_t}{2\pi}\,
\left\{
T_1^\mu\,(\bar \ell\gamma_\mu \ell)+T_2^\mu\,(\bar \ell\gamma_\mu\gamma_5 \ell)
\right\},
\en
where the quantities $T_i^\mu$ are expressed through the form
factors and the Wilson coefficients in the following manner:

\be
\label{amp_pv}
T_i^\mu =  T_i^{\mu\nu}\,\epsilon^\dagger_{2\nu}\,, 
\qquad (i=1,2)\,,
\en

\bea 
T_i^{\mu\nu} &=& \frac{1}{m_1+m_2} \left\{
- Pq\,g^{\mu\nu}\,A_0^{(i)} + P^\mu P^\nu\,A_+^{(i)} + q^\mu P^\nu\, A_-^{(i)} 
+ i\varepsilon^{\mu\nu\alpha\beta}P_\alpha q_\beta\,V^{(i)}  
\right\}\,, 
\nn[2ex]
V^{(1)} &=&   C_9^{\rm eff}\,V  + C_7^{\rm eff}\,g \,\frac{2\bar m_b(m_1+m_2)}{q^2}\,,
\nn
A_0^{(1)} &=& C_9^{\rm eff}\,A_0 
+ C_7^{\rm eff}\,a_0\,\frac{2\bar m_b(m_1+m_2)}{q^2}\,,
\nn
A_+^{(1)} &=& C_9^{\rm eff}\,A_+ 
+ C_7^{\rm eff}\,a_+\,\frac{2\bar m_b(m_1+m_2)}{q^2}\,,
\nn
A_-^{(1)} &=& C_9^{\rm eff}\,A_- 
+ C_7^{\rm eff}\,(a_0-a_+)\,\frac{2\bar m_b(m_1+m_2)}{q^2}\,\frac{Pq}{q^2}\,,
\nn[1.5ex]
V^{(2)}   &=& C_{10}\,V, \qquad A_0^{(2)} = C_{10}\,A_0,\qquad
A_\pm^{(2)} = C_{10}\,A_\pm.
\ena

\begin{figure}[ht]
\begin{center}
\epsfig{figure=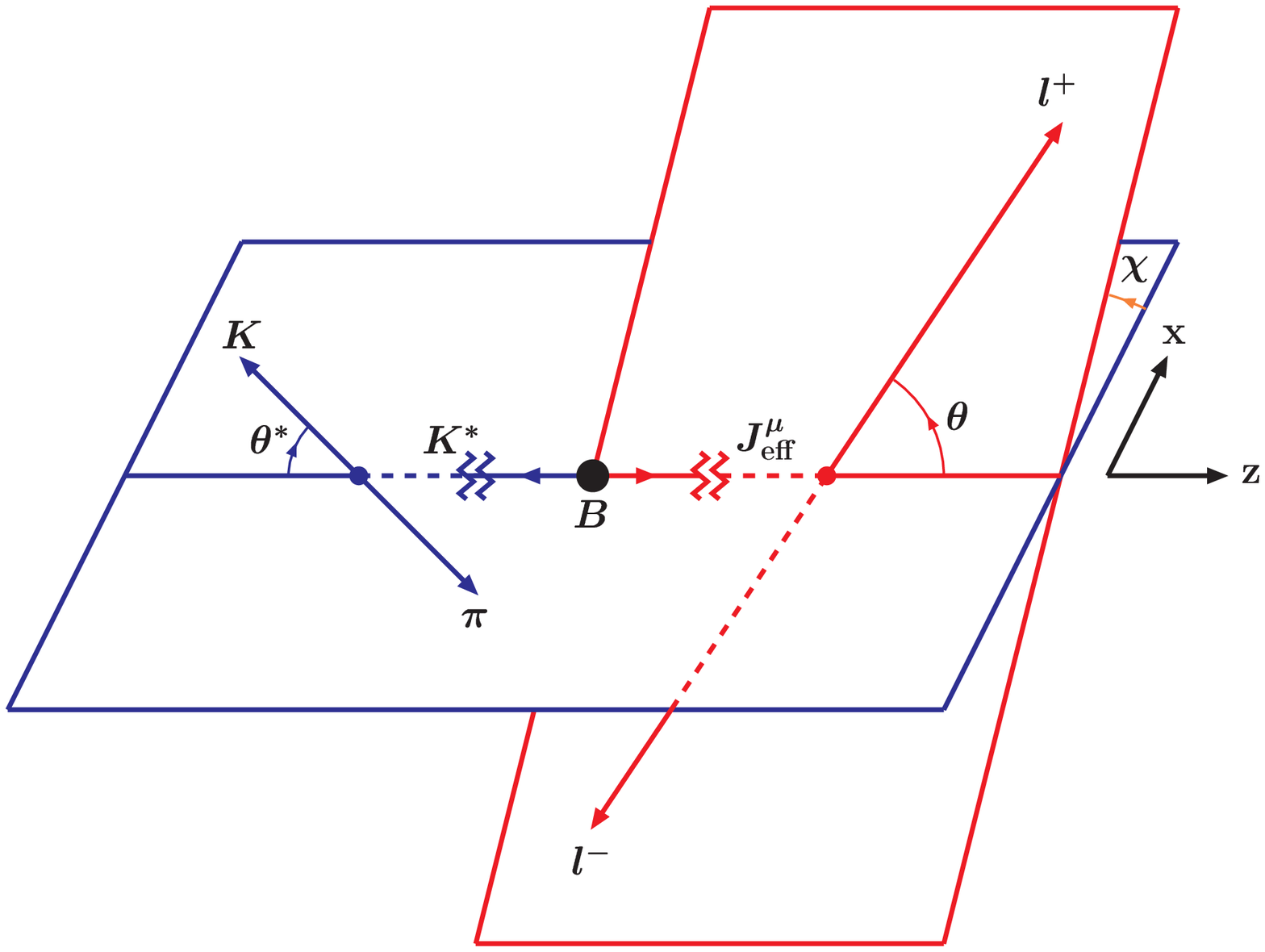,scale=0.7}
\caption{Definition of the angles $\theta$, $\theta^\ast$ and $\chi$ in
the cascade decay $B\to K^\ast(\to K\pi)\bar \ell \ell$.}
\label{fig:bkangl}
\end{center}
\end{figure}

Let us consider the decay distribution differential in the momentum transfer 
squared $q^2$ and in the polar angle. The latter is defined
as the angle between $\vec q=\vec p_1-\vec p_2$ and $\vec k_1$
($\ell^+\ell^-$ rest frame) as shown in Fig.~\ref{fig:bkangl}. 
By using the notation from Ref.~\cite{Faessler:2002ut} and correcting
some obvious typos, one has
\bea
\frac{d^2\Gamma}{dq^2 d\cos\theta} &=& 
\frac{|{\bf p_2}| \, \beta_\ell}{(2\pi)^3\,4\,m_1^2}
\cdot\frac{1}{8}\sum\limits_{\rm pol}|M|^2
=\frac{G^2_F}{(2\pi)^3}\, 
\biggl(\frac{\alpha|\lambda_t|}{2\,\pi}\biggr)^2
\frac{|{\bf p_2}|\,\beta_\ell }{8 m_1^2}
\label{eq:2-fold-dis}\\
&\times& \frac{1}{8}
\biggl\{H^{\mu\nu}_{11}\cdot
{\rm tr} [\gamma_\mu\,(\not\! k_1-m_\ell)\,\gamma_\nu\,
(\not\! k_2+m_\ell) ]
\nn
&+& H^{\mu\nu}_{22}\cdot {\rm tr} [\gamma_\mu\gamma_5 \,
(\not\! k_1-m_\ell)\,\gamma_\nu\gamma_5\,(\not\! k_2+m_\ell) ]
\nn
&-&\,H^{\mu\nu}_{12}\cdot {\rm tr} [\gamma_\mu\,(\not\! k_1-m_\ell)\,
              \gamma_\nu\gamma_5\,(\not\! k_2+m_\ell) ]
\nn
&-& H^{\mu\nu}_{21}\cdot
{\rm tr} [\gamma_\mu\gamma_5\,(\not\! k_1-m_\ell)\,
              \gamma_\nu\,(\not\! k_2+m_\ell) ]
\biggr\}
\nn[2ex]
&=&
\frac{G^2_F}{(2\pi)^3}\,\biggl(\frac{\alpha|\lambda_t|}{2\,\pi}\biggr)^2
\frac{|{\bf p_2}|\,\beta_\ell } {8 m_1^2}
\cdot\frac{1}{2} \biggl\{
 L^{(1)}_{\mu\nu}\cdot (H^{\mu\nu}_{11}+H^{\mu\nu}_{22})
\nn
&-& \frac{1}{2}\,L^{(2)}_{\mu\nu}\cdot 
(q^2\,H^{\mu\nu}_{11}+  (q^2 - 4m_\ell^2)\,H^{\mu\nu}_{22})
+L^{(3)}_{\mu\nu}\cdot (H^{\mu\nu}_{12}+H^{\mu\nu}_{21})
\biggr\},
\nonumber
\ena
where  $|{\bf p_2}|=\lambda^{1/2}(m_1^2,m_2^2,q^2)/2m_1$ is
the momentum of the $K^\ast$-meson given in the $B$-rest frame
and $\beta_\ell=\sqrt{1-4m_\ell^2/q^2}$. 
We have introduced lepton and hadron tensors as
\bea
L^{(1)}_{\mu\nu} &=& k_{1\mu} k_{2\nu}+ k_{2\mu} k_{1\nu}\,,
\qquad
L^{(2)}_{\mu\nu}= g_{\mu\nu}\,,
\qquad
L^{(3)}_{\mu\nu}= i\varepsilon_{\mu\nu\alpha\beta}k_1^\alpha k_2^\beta\,,
\label{eq:lept_tensor}\\
H_{ij}^{\mu\nu} & = & T_i^\mu\,T_j^{\dagger\nu}\,.
\label{eq:had_tensor}
\ena
We use the following convention for the $\gamma_5$-matrix and the Levy-Civita
tensor in Minkowski space: 
\bea
&&
\gamma^0 = \left(\begin{array}{lr}
                           I & 0 \\
                           0 & -I
                    \end{array}\right)\,, \qquad
\gamma^k = \left(\begin{array}{lr}
                               0     & \sigma_k \\
                           -\sigma_k & 0
                    \end{array}\right)\,, \quad
\gamma^5 = \gamma_5 = i\,\gamma^0\gamma^1\gamma^2\gamma^3
=  \left(\begin{array}{lr}
                           0 & I \\
                           I & 0
                    \end{array}\right)\,, 
\label{eq:Levy-Civita}\\[1.3ex]
&&
\Tr\left(\gamma_5\gamma^\mu\gamma^\nu\gamma^\alpha\gamma^\beta \right)
= 4\,i\,\varepsilon^{\mu\nu\alpha\beta}\,,
\qquad
\Tr\left(\gamma_5\gamma_\mu\gamma_\nu\gamma_\alpha\gamma_\beta \right)
= 4\,i\, \varepsilon_{\mu\nu\alpha\beta}\,,
\qquad
\varepsilon_{0123} = -\varepsilon^{0123} = +1.
\nonumber
\ena

The Lorentz contractions in Eq.~(\ref{eq:2-fold-dis}) can be  evaluated in 
terms of helicity amplitudes as described  in 
\cite{Korner:1989qb,Faessler:2002ut}.
First, we define an orthonormal and complete helicity basis
$\epsilon^\mu(m)$ with the three spin 1 components  orthogonal to
the momentum transfer $q^\mu$, i.e. $\epsilon^\mu(m) q_\mu=0$ for 
$m=\pm,0$, and the spin 0 (time)-component $m=t$ with
$\epsilon^\mu(t)= q^\mu/\sqrt{q^2}$.   
The orthonormality and completeness properties read 
\bea
\epsilon^\dagger_\mu(m)\epsilon^\mu(n) &=& g_{mn}\,, \quad (m,n=t,\pm,0), 
\quad \text{orthonormality}\,,
\nn[1.5ex]
\epsilon_\mu(m)\epsilon^{\dagger}_{\nu}(n)g_{mn} &=& g_{\mu\nu} \,,\quad 
\phantom{(m,n=t,\pm,0),} \quad\,\,\, \text{completeness},
\label{eq:orth-compl}
\ena
with $g_{mn}={\rm diag}\,(\,+\,,\,\,-\,,\,\,-\,,\,\,-\,)$.
We include the time component polarization vector $\epsilon^\mu(t)$
in the set because we want to discuss lepton mass effects.

Using the completeness property one can rewrite the contraction
of the lepton and hadron tensors in Eq.~(\ref{eq:2-fold-dis}) as follows
\bea
L^{(k)\mu\nu}H_{\mu\nu}^{ij} &=& 
L_{\mu'\nu'}^{(k)}\epsilon^{\mu'}(m)\epsilon^{\dagger\mu}(m')g_{mm'}
\epsilon^{\dagger \nu'}(n)\epsilon^{\nu}(n')g_{nn'}H_{\mu\nu}^{ij}
\nn
&=& L^{(k)}_{mn} g_{mm'} g_{nn'} H^{ij}_{m'n'},
\label{eq:contraction}
\ena
where we have introduced the lepton and hadron tensors in the space
of the helicity components
\be
L^{(k)}_{mn} = \epsilon^\mu(m)\epsilon^{\dagger \nu}(n)L^{(k)}_{\mu\nu},
\qquad
H^{ij}_{mn} = \epsilon^{\dagger\mu}(m)\epsilon^\nu(n)H^{ij}_{\mu\nu}\, .
\label{eq:hel_tensors}
\en
The two tensors can be evaluated in two different
Lorentz systems. The lepton tensors $L^{(k)}_{mn}$ will be evaluated
in the $\bar \ell\ell$-CM system whereas the hadron tensors $H^{ij}_{mn}$
will be evaluated in the $B$ rest system.

In the $B$ rest frame the momenta and polarization vectors
can be written as
\be
\begin{aligned}
p^\mu_1 &= (\,m_1\,,\,\,0,\,\,0,\,\,0\,), & \qquad
\epsilon^\mu(t) & =
\frac{1}{\sqrt{q^2}}(\,q_0\,,\,\,0\,,\,\,0\,,\,\,|{\bf p_2}|\,),
\nn
p^\mu_2 &= (\,E_2\,,\,\,0\,,\,\,0\,,\,\,-|{\bf p_2}|\,), & \qquad
\epsilon^\mu(\pm) &= 
\frac{1}{\sqrt{2}}(\,0\,,\,\,\mp 1\,,\,\,-i\,,\,\,0\,), &\qquad
\nn
q^\mu   &= (\,q_0\,,\,\,0\,,\,\,0\,,\,\,+|{\bf p_2}|\,), & \qquad
\epsilon^\mu(0) &=
\frac{1}{\sqrt{q^2}}(\,|{\bf p_2}|\,,\,\,0\,,\,\,0\,,\,\,q_0\,),
\end{aligned}
\label{eq:B-rest}
\en
where $E_2 = (m_1^2+m_2^2-q^2)/2 m_1$ and $q_0=(m_1^2-m_2^2+q^2)/2 m_1$.
Using this basis one can express the components of the hadronic
tensors through the invariant form factors defined in  
Eqs.~(\ref{eq:PV}) and (\ref{eq:PVT}). One has
\bea 
H^{ij}_{mn} &=&  
\epsilon^{\dagger \mu}(m) \epsilon^{ \nu}(n)H^{ij}_{\mu\nu}
=
\epsilon^{\dagger \mu}(m) \epsilon^{ \nu}(n) 
T^i_{\mu\alpha} 
\left(-g^{\alpha\beta}+\frac{p_2^\alpha p_2^\beta}{m_2^2}\right)
T^{\dagger j}_{\beta\nu}
\nn
&=&
\epsilon^{\dagger \mu}(m) \epsilon^{ \nu}(n) 
T^i_{\mu\alpha}
\epsilon_2^{\dagger\alpha}(r)\epsilon_2^{\beta}(s)\delta_{rs}
T^{\dagger j}_{\beta\nu}
\nn
&=&
\epsilon^{\dagger \mu}(m)\epsilon_2^{\dagger\alpha}(r)
T^i_{\mu\alpha} \cdot
\left(\epsilon^{\dagger \nu}(n)\epsilon_2^{\dagger\beta}(s)T^j_{\nu\beta}
\right)^\dagger\delta_{rs}
\equiv H^i_m H^{\dagger \,j}_n.
\label{eq:hel_vv_def}
\ena 
From angular momentum conservation one has 
$r=m$ and $s=n$ for $m,n=\pm,0$ and $r,s=0$ for $m,n=t$.
The helicity components $\epsilon_2(m)$ $(m=\pm,0)$ of the polarization 
vector of the $K^\ast$ read
\be
\epsilon^\mu_2(\pm) = 
\frac{1}{\sqrt{2}}(0\,,\,\,\pm 1\,,\,\,-i\,,\,\,0\,)\,,
\qquad
\epsilon^\mu_2(0) = 
\frac{1}{m_2}(|{\bf p_2}|\,,\,\,0\,,\,\,0\,,\,\,-E_2\,)\,.
\label{eq:vect_pol}
\en
Then one obtains the non-zero components of the hadron tensors 
\bea
H^i_{t0} &=& 
\epsilon^{\dagger \mu}(t)\epsilon_2^{\dagger \alpha}(0)T^i_{\mu\alpha}
\,=\,
\frac{1}{m_1+m_2}\frac{m_1\,|{\bf p_2}|}{m_2\sqrt{q^2}}
\left(Pq\,(-A^i_0+A^i_+)+q^2 A^i_-\right),
\nn[1.2ex]
H^i_{\pm1\pm1} &=& 
\epsilon^{\dagger \mu}(\pm)\epsilon_2^{\dagger \alpha}(\pm)
T^i_{\mu\alpha}
\,=\,
\frac{1}{m_1+m_2}\left(-Pq\, A^i_0\pm 2\,m_1\,|{\bf p_2}|\, V^i \right),
\nn[1.2ex]
H^i_{00} &=&  
\epsilon^{\dagger \mu}(0)\epsilon_2^{\dagger \alpha}(0)T^i_{\mu\alpha} =
\nn[1.2ex]
&=&
\frac{1}{m_1+m_2}\frac{1}{2\,m_2\sqrt{q^2}} 
\left(-Pq\,(m_1^2 - m_2^2 - q^2)\, A^i_0 + 4\,m_1^2\,|{\bf p_2}|^2\, A^i_+\right).
\label{eq:hel_vv}
\ena
The lepton tensors $L^{(k)}_{mn}$ are evaluated 
in the $\bar \ell\ell$-CM system $\vec k_1+\vec k_2=0$.
\be
\begin{array}{lr}
q^\mu   = (\,\sqrt{q^2}\,,\,\,0\,,\,\,0\,,\,\,0\,)\,,\qquad
&
\qquad
k^\mu_1 = 
(\,E_1\,,\,\,+|{\bf k_1}|\sin\theta\,,\,\, 0\,,\,\,+|{\bf k_1}|\cos\theta\,)\,,
\nn[1.2ex]
\qquad
& k^\mu_2 = 
(\,E_1\,,\,\,-|{\bf k_1}|\sin\theta\,,\,\,0 \,,\,\,-|{\bf k_1}|\cos\theta\,)\,,
\end{array}
\label{eq:kpi_basis}
\en
with $E_1=\sqrt{q^2}/2$ and $|{\bf k_1}|=\sqrt{q^2-4 m_\ell^2}/2$.
The longitudinal and time component of polarization vectors 
in the $\bar \ell \ell$ rest frame 
are given by  
\[\epsilon(t)=(1,0,0,0),\quad 
\epsilon^\mu(\pm) = \frac{1}{\sqrt{2}}(0,\mp 1,-i,0),\quad
\epsilon^\mu(0)=(0,0,0,1).
\]
The differential $(q^2,\cos\theta)$ distribution finally reads
\bea
\frac{d\Gamma(B\to K^\ast\bar \ell \ell)}{dq^2d\cos\theta} &=&\,
 \frac{G^2_F}{(2\pi)^3} 
\left(\frac{\alpha|\lambda_t|}{2\pi}\right)^2
\frac{|{\bf p_2}|\,q^2\,\beta_\ell}{32\,m_1^2}
\nn
&\times&
\Big\{
 (1+\cos^2\theta)\cdot\frac12\left( {\cal H}^{11}_U + {\cal H}^{22}_U \right )
+ \sin^2\theta\,\left( {\cal H}^{11}_L + {\cal H}^{22}_L \right )
- 2\,\beta_\ell \cos\theta\cdot {\cal H}^{12}_P
\nn
&& +\, \delta_{\ell\ell}\,\left[\, \sin^2\theta\cdot  {\cal H}^{11}_U 
                 - (1+\cos^2\theta)\cdot  {\cal H}^{22}_U
\right.
\nn 
&&
\left.
                 +\,2\,\cos^2\theta\cdot  {\cal H}^{11}_L 
                 -2\,\sin^2\theta\cdot  {\cal H}^{22}_L 
                 +\,2\,  {\cal H}^{22}_S 
                 \right]
\Big\}, 
\label{eq:distr2}
\ena
%--------------- insertion -------------------------
where the helicity flip suppression factor $\delta_{\ell\ell}$
is given by 
\[ 
\delta_{\ell\ell} = \frac{2m^2_\ell}{q^2}.\]
After integrating over $\cos\theta$ one obtains 
\bea
&&
\frac{d\Gamma(B\to K^\ast\bar \ell \ell)}{dq^2} =\,
 \frac{G^2_F}{(2\pi)^3}\,
\left(\frac{\alpha|\lambda_t|}{2\pi}\right)^2
\frac{|{\bf p_2}|\,q^2\,\beta_\ell}{12\,m_1^2}
{\cal H}_{\rm tot}\,,
\nn[1.2ex]
&&
{\cal H}_{\rm tot} = 
\frac12\left(  {\cal H}^{11}_U + {\cal H}^{22}_U 
             + {\cal H}^{11}_L + {\cal H}^{22}_L \right )
+ \delta_{\ell\ell}\,
\left[\,\frac12 {\cal H}^{11}_U - {\cal H}^{22}_U
       + \frac12 {\cal H}^{11}_L - {\cal H}^{22}_L + \frac32\, {\cal H}^{22}_S
\right].
\label{eq:distr1}
\ena
The relevant bilinear combinations of the helicity amplitudes
are defined in Table~\ref{tab:helicity}. Note that we drop
a factor of ``3'' in the notation of ${\cal H}_S$ and ${\cal H}_{IS}$
compared with our paper \cite{Faessler:2002ut}.
\begin{table}[hb] 
\begin{center}
\caption{ Definition of helicity structure functions and their parity 
properties.}
\def\arraystretch{1}
\begin{tabular}{ll}
\hline
parity-conserving (p.c.) \qquad & \qquad  parity-violating (p.v.)  \\
\hline
${\cal H}^{ij}_U   =  \re\left(H^i_{+1 +1} H^{\dagger\,j}_{+1 +1}\right) 
               + \re\left(H^i_{-1 -1} H^{\dagger\,j}_{-1 -1}\right)$   
\qquad &  \qquad
${\cal H}^{ij}_P   =   \re\left(H^i_{+1 +1} H^{\dagger\,j}_{+1 +1}\right) 
               - \re\left(H^i_{-1 -1} H^{\dagger\,j}_{-1 -1}\right)$   
\\
${\cal H}^{ij}_{IU} =  \im\left(H^i_{+1 +1} H^{\dagger\,j}_{+1 +1}\right) 
               + \im\left(H^i_{-1 -1} H^{\dagger\,j}_{-1 -1}\right)$   
\qquad &  \qquad
${\cal H}^{ij}_{IP}   =   \im\left(H^i_{+1 +1} H^{\dagger\,j}_{+1 +1}\right) 
               -  \im\left(H^i_{-1 -1} H^{\dagger\,j}_{-1 -1}\right)$   
\\
${\cal H}^{ij}_L   =  \re\left(H^i_{0 0} H^{\dagger\,j}_{0 0}\right) $
\qquad &  \qquad
${\cal H}^{ij}_A   = \frac12\Big[ \re\left(H^i_{+1 +1} H^{\dagger\,j}_{0 0}\right) 
                   - \re\left(H^i_{-1 -1} H^{\dagger\,j}_{0 0}\right)\Big]$  
\\
${\cal H}^{ij}_{IL}   =  \im\left(H^i_{0 0} H^{\dagger\,j}_{0 0}\right) $
\qquad &  \qquad
${\cal H}^{ij}_{IA}   = \frac12\Big[ \im\left(H^i_{+1 +1} H^{\dagger\,j}_{0 0}\right) 
                   - \im\left(H^i_{-1 -1} H^{\dagger\,j}_{0 0}\right)\Big]$  
\\
${\cal H}^{ij}_{T} =\re\left( H^i_{+1+1}H_{-1-1}^{\dagger\,j}\right)$
\qquad &  \qquad
${\cal H}^{ij}_{SA} = 
\frac12 \,\Big[  \re\left(  H^i_{+1 +1}  H_{0\, t}^{\dagger\,j} \right) 
                 -\re\left( H^i_{-1 -1} H_{0\,t}^{\dagger\,j} \right)\Big]$
\\ 
${\cal H}^{ij}_{IT} =\im\left( H^i_{+1+1}H_{-1-1}^{\dagger\,j}\right)$
\qquad &  \qquad
${\cal H}^{ij}_{ISA} = 
 \frac12 \,\Big[ \im\left(  H^i_{+1 +1}  H_{0\, t}^{\dagger\,j} \right) 
                -\im\left(  H^i_{-1 -1} H_{0\,t}^{\dagger\,j} \right)\Big]$
\\ 
${\cal H}^{ij}_{I}  =  
 \frac12 \,\Big[  \re\left( H^i_{+1 +1} H_{0\,0}^{\dagger\,j} \right) 
                + \re\left( H^i_{-1 -1} H_{0\,0}^{\dagger\,j} \right)\Big]$ 
 \qquad &  \qquad 
\\
${\cal H}^{ij}_{II}  =  
 \frac12 \,\Big[  \im\left( H^i_{+1 +1} H_{0\,0}^{\dagger\,j} \right) 
                + \im\left( H^i_{-1 -1} H_{0\,0}^{\dagger\,j} \right)\Big] $ 
 \qquad &  \qquad 
\\
${\cal H}^{ij}_S    = \re\left(H^i_{0 t}H^{\dagger\,j}_{0 t}\right)$ 
\qquad &  \qquad  
\\
${\cal H}^{ij}_{IS}    = \im\left(H^i_{0 t}H^{\dagger\,j}_{0 t}\right)$ 
\qquad &  \qquad  
\\
 ${\cal H}^{ij}_{ST} =  
 \frac12\,\Big[ \re\left( H^i_{+1 +1} H_{0\,t}^{\dagger\,j} \right)
               +\re\left( H^i_{-1 -1} H_{0\,t}^{\dagger\,j} \right)\Big]$
 \qquad &  \qquad  
\\ 
${\cal H}^{ij}_{IST} =  
 \frac12 \,\Big[ \im\left( H^i_{+1 +1} H_{0\,t}^{\dagger\,j} \right)
                +\im\left( H^i_{-1 -1} H_{0\,t}^{\dagger\,j} \right)\Big]$
 \qquad &  \qquad  
\\ 
${\cal H}^{ij}_{SL} =  \re\left(  H^i_{0\,0} H_{0\,t}^{\dagger\,j} \right) $
 \qquad &  \qquad
 \\
${\cal H}^{ij}_{ISL} = \im\left(  H^i_{0\,0} H_{0\,t}^{\dagger\,j} \right) $
 \qquad &  \qquad
 \\[2ex]
\hline
\end{tabular}
\label{tab:helicity}
\end{center}
\end{table}

\section{Four-fold distribution 
in cascade decay  $B\to K^\ast(\to K\pi) \bar \ell\ell$}
\label{sec:four-fold}

The lepton-hadron correlation function $L_{\mu\nu}H^{\mu\nu}$ reveals 
even more structure when one uses the cascade decay 
$B\to K^\ast(\to K\pi) \bar \ell\ell$ to analyze the polarization of the 
$K^\ast$. The hadron tensor now reads \cite{Faessler:2002ut}
\be
H^{ij}_{\mu\nu}=T^i_{\mu\alpha}(T^j_{\nu\beta})^\dagger
\frac{3}{2\,|{\bf p_3}|}{\rm Br}(K^\ast\to K\pi)p_{3\alpha'}p_{3\beta'}
S^{\alpha\alpha'}(p_2)S^{\beta\beta'}(p_2),
\label{eq:had-ten-4}
\en
where 
$S^{\alpha\alpha'}(p_2)=-g^{\alpha\alpha'}+p_2^\alpha 
p_2^{\alpha'}/m_2^2$ 
is the standard spin 1 tensor,
$p_2=p_3+p_4$, $p_3^2=m_K^2$, $p_4^2=m_\pi^2$, and  $p_3$ and $p_4$
are the momenta of the $K$ and the $\pi$, respectively. 
The relative configuration of the ($K,\pi$)- and ($\bar \ell \ell$)-planes
is shown in Fig.~\ref{fig:bkangl}. 
The branching ratio ${\rm Br}(K^\ast\to K\pi)\approx 1$ so we drop it off
in what follows. 

In the rest frame of the $K^\ast$ one has 
\bea
p^\mu_2 &=& (m_2,\vec{0}),
\nn
p^\mu_3 &=& (E_3 , +|{\bf p_3}|\sin\theta^\ast , 0 , -|{\bf p_3}|\cos\theta^\ast),
\nn
p^\mu_4 &=& (E_4 , -|{\bf p_3}|\sin\theta^\ast , 0 ,+|{\bf p_3}|\,\cos\theta^\ast), 
\label{eq:daughter-frame}
\ena
where $ |{\bf p_3}| = \lambda^{1/2}(m_2^2,m_3^2,m_4^2)/(2\,m_2)$.
According to Eq.~(\ref{eq:vect_pol}) the rest frame polarization vectors
of the $K^\ast$ are given by
\be
\epsilon^\mu_2(\pm) = \frac{1}{\sqrt{2}}(\,0\,,\,\,\pm 1\,,\,\,
-i\,,\,\,0\,)\,,\qquad
\epsilon^\mu_2(0) = (\,0\,,\,\,0\,,\,\,0\,,\,\,-1)\,.
\label{eq:casc_pol}
\en
The spin 1 tensor $S^{\alpha\alpha'}(p_2)$ is then written as 
\be
S^{\alpha\alpha'}(p_2)=-g^{\alpha\alpha'} + \frac{p_2^\alpha p_2^{\alpha'}}{m_2^2}
=\sum\limits_{m=\pm,0}\epsilon_2^\alpha(m)\epsilon_2^{\dagger\alpha'}(m) \, . 
\label{eq:S=1-tensor}
\en
Following basically the same trick as in Eq.~(\ref{eq:contraction})
the contraction of the lepton and hadron tensors may be
written through helicity components. Finally, one obtains  
the full four-fold angular decay distribution 
\bea
&&\frac{d\Gamma(B\to K^\ast(\to K\pi)\bar \ell \ell)}
     {dq^2\,d\cos\theta\,d(\chi/2\pi)\,d\cos\theta^\ast} =
\frac{G^2_F}{(2\pi)^3}\,
\left(\frac{\alpha|\lambda_t|}{2\pi}\right)^2
\frac{|{\bf p_2}|\,q^2\,\beta_\ell}{12\,m_1^2}\,W(\theta^\ast,\theta,\chi),
\nn[1.2ex]
W(\theta^\ast,\theta,\chi) &=&
\frac{9}{64}\,(1+\cos^2\theta)\sin^2\theta^\ast
\left( {\cal H}^{11}_U +  {\cal H}^{22}_U \right)
+
\frac{9}{16} \sin^2\theta\cos^2\theta^\ast
\left( {\cal H}^{11}_L +  {\cal H}^{22}_L \right)
\nn
&-& \frac{9}{32}\,\sin^2\theta\sin^2\theta^\ast\cos 2\chi
\left( {\cal H}^{11}_T +  {\cal H}^{22}_T \right)
+
\frac{9}{32}\,\sin 2\theta\sin 2\theta^\ast \cos\chi
\left( {\cal H}^{11}_I +  {\cal H}^{22}_I \right)
\nn
&+& \beta_\ell\,
\left[
-\frac{9}{16}\,\cos\theta\sin^2\theta^\ast {\cal H}_P^{12}
\right.
\nn
&&
\left.
-\frac{9}{16}\,\sin\theta\sin 2\theta^\ast\cos\chi
\left( {\cal H}^{12}_A +  {\cal H}^{21}_A \right)
+\frac{9}{16}\,\sin\theta\sin 2\theta^\ast\sin\chi
\left( {\cal H}^{12}_{II} +  {\cal H}^{21}_{II} \right)
\right]
\nn
&-& \frac{9}{32}\,\sin 2\theta\sin 2\theta^\ast\sin\chi 
\left( {\cal H}^{11}_{IA} +  {\cal H}^{22}_{IA} \right)
+
\frac{9}{32}\,\sin^2\theta\sin^2\theta^\ast\sin 2\chi
\left( {\cal H}^{11}_{IT} +  {\cal H}^{22}_{IT} \right)
\nn[1.2ex]
&+&\delta_{\ell\ell}
\Big\{
 \frac{9}{32}\,\sin^2\theta\sin^2\theta^\ast {\cal H}^{11}_U
- 
 \frac{9}{32}\,(1+\cos^2\theta)\sin^2\theta^\ast {\cal H}^{22}_U
\nn[1.2ex]
&& +\frac{9}{8}\,\cos^2\theta\cos^2\theta^\ast {\cal H}^{11}_L
   -\frac{9}{8}\,\sin^2\theta\cos^2\theta^\ast  {\cal H}^{22}_L
\nn
&&
+\frac{9}{16}\,\sin^2\theta\sin^2\theta^\ast \cos 2\chi
\left( {\cal H}^{11}_{T} +  {\cal H}^{22}_{T} \right)
\nn[1.2ex]
&&
-\frac{9}{16}\,\sin 2\theta\sin 2\theta^\ast \cos\chi
\left( {\cal H}^{11}_{I} +  {\cal H}^{22}_{I} \right)
+
\frac{9}{8}\,\cos^2\theta^\ast {\cal H}^{22}_{S} 
\nn
&&
+\frac{9}{16}\,\sin 2\theta\sin 2\theta^\ast\sin\chi
\left( {\cal H}^{11}_{IA} +  {\cal H}^{22}_{IA} \right)
\nn
&&
- \frac{9}{16}\,\sin^2\theta\sin^2\theta^\ast\sin 2\chi
\left( {\cal H}^{11}_{IT} +  {\cal H}^{22}_{IT} \right)
\Big\}.
\label{eq:distr4}
\ena 

The differential angular decay distribution in 
the Ref.~(\cite{Kruger:2005ep}) is expressed via
the transversality  amplitudes $A_\perp$, $A_\parallel$, $A_0$ and $A_t$.
They are related to our helicity amplitudes as
\bea
A^{L,R}_\perp &=& N \,\frac{1}{\sqrt{2}}
\left[(H^{(1)}_{+1+1} - H^{(1)}_{-1-1}) \mp (H^{(2)}_{+1+1} - H^{(2)}_{-1-1}) \right]\,,
\nn[1.2ex]
A^{L,R}_\parallel &=& N \,\frac{1}{\sqrt{2}}
\left[ (H^{(1)}_{+1+1} + H^{(1)}_{-1-1}) \mp (H^{(2)}_{+1+1} + H^{(2)}_{-1-1}) \right]\,,
\nn[1.2ex]
A^{L,R}_0 &=& N\,\left( H^{(1)}_{00} \mp   H^{(2)}_{00} \right)\,,
\nn[1.2ex]
A_t &=& -2\,N\,H^{(2)}_{0t},
\label{eq:eq:KGvsOur}
\ena
where the overall factor is given by
\[
N = \Big[ \frac14 \frac{G_F^2}{(2\pi)^3} 
                   \left( \frac{\alpha |\lambda_t|}{2\pi} \right)^2
                  \frac{ {\bf|p_2|} q^2 \beta_\ell }{12 m_1^2 } 
     \Big]^{\frac12} \,.
\]

One can check that the four-fold angular distribution in Eq.(\ref{eq:distr4})
agrees with those obtained in \cite{Kruger:2005ep} Eq.(3.1).

%----------------- angle distribution from Matias 1202.4266 ---

The four-body distribution of the cascade decay 
$B_d^0\to K^{\ast\,0}(\to K\pi)\,\ell^+\ell^-$ is now often written 
in the form as given in Ref.~\cite{Matias:2012xw}
\eqa{\label{dist}
\frac{d^4\Gamma}{dq^2\,d\!\cos\theta^\ast\,d\!\cos\theta\,d\chi}&=&\frac9{32\pi}
\bigg[
   J_{1s} \sin^2\theta^\ast + J_{1c} \cos^2\theta^\ast 
+ (J_{2s} \sin^2\theta^\ast + J_{2c} \cos^2\theta^\ast) \cos 2\theta
\nn[1.2ex]
&&\hspace{-2.7cm}
+ J_3 \sin^2\theta^\ast \sin^2\theta \cos 2\chi 
+ J_4 \sin 2\theta^\ast \sin 2\theta \cos\chi  
+ J_5 \sin 2\theta^\ast \sin\theta \cos\chi 
\nn[1.5mm]
&&\hspace{-2.7cm}
+ (J_{6s} \sin^2\theta^\ast +  {J_{6c} \cos^2\theta^\ast})  \cos\theta    
+ J_7 \sin 2\theta^\ast \sin\theta \sin\chi  
+ J_8 \sin 2\theta^\ast \sin 2\theta \sin\chi 
\nn[1.5mm]
&&\hspace{-2.7cm}+ J_9 \sin^2\theta^\ast \sin^2\theta \sin 2\chi \bigg]\, ,
}
where the expressions for the coefficients $J_i$ are written as
\eqa{
J_{1s}  & = & 
\frac{(2+\beta_\ell^2)}{4} 
 \left[|\apeL|^2 + |\apaL|^2 +|\apeR|^2 + |\apaR|^2 \right]
+ \frac{4 m_\ell^2}{q^2} \re\left(\apeL\apeR^* + \apaL\apaR^*\right)\,,
\nn[1mm]
J_{1c}  & = &  |\azeL|^2 +|\azeR|^2  
+ \frac{4m_\ell^2}{q^2} 
\left[|A_t|^2 + 2\re(\azeL^{}\azeR^*) \right] + \beta_\ell^2\, |A_S|^2 \,,
\nn[1mm]
J_{2s} & = & \frac{ \beta_\ell^2}{4}
\left[ |\apeL|^2+ |\apaL|^2 + |\apeR|^2+ |\apaR|^2\right],
\hspace{0.92cm}    
J_{2c}  = - \beta_\ell^2\left[|\azeL|^2 + |\azeR|^2 \right]\,,
\nn[1mm]
J_3 & = & \frac{1}{2}\beta_\ell^2
\left[ |\apeL|^2 - |\apaL|^2  + |\apeR|^2 - |\apaR|^2\right],
\qquad   
J_4  = \frac{1}{\sqrt{2}}\beta_\ell^2
\left[\re (\azeL\apaL^* + \azeR\apaR^* )\right],
\nn[1mm]
J_5 & = & \sqrt{2}\beta_\ell\,\Big[\re(\azeL\apeL^* - \azeR\apeR^* ) 
- \frac{m_\ell}{\sqrt{q^2}}\,\re(\apaL A_S^*+ \apaR^* A_S) \Big]\,,
\nn[1mm]
J_{6s} & = &  2\beta_\ell\left[\re (\apaL\apeL^* - \apaR\apeR^*) \right]\,,
\hspace{2.25cm} 
J_{6c} = 4\beta_\ell\, \frac{m_\ell}{\sqrt{q^2}}\,\re (\azeL A_S^*+ \azeR^* A_S)
\,,\nn[1mm]
J_7 & = & \sqrt{2} \beta_\ell\, \Big[\im (\azeL\apaL^* - \azeR\apaR^* ) 
+ \frac{m_\ell}{\sqrt{q^2}}\, \im (\apeL A_S^* - \apeR^* A_S)) \Big]\,,
\nn[1mm]
J_8 & = & \frac{1}{\sqrt{2}}\beta_\ell^2
\left[\im(\azeL\apeL^* + \azeR\apeR^*)\right]\,,
\hspace{1.9cm} 
J_9 = \beta_\ell^2\left[\im (\apaL^{*}\apeL + \apaR^{*}\apeR)\right] \,.
\label{Js}}\\

We do not consider here the CP-violating observables and scalar
contributions $A_S\equiv~0$.  Following Ref.~\cite{Matias:2012xw}
we choose, first, three natural observables: the differential 
branching fraction $d{\cal B}/dq^2$, the forward-backward asymmetry 
$A_{\rm FB}$ and the longitudinal polarization 
\eqa{
\frac{d\Gamma}{dq^2} &=& \int d\!\cos\theta\, d\!\cos\theta^\ast d\chi\,
\frac{d^4\Gamma}{dq^2\,d\!\cos\theta^\ast\,d\!\cos\theta\,d\chi}
= \dfrac{1}{4} \left(3 J_{1c} + 6 J_{1s} - J_{2c} -2 J_{2s}\right)
\nn[2ex]
&=& \frac{G^2_F}{(2\pi)^3}\,
\left(\frac{\alpha|\lambda_t|}{2\pi}\right)^2
\frac{|{\bf p_2}|\,q^2\,\beta_\ell}{12\,m_1^2}
{\cal H}_{\rm tot}\,, \qquad 
\frac{d{\cal B}}{dq^2} = \frac{1}{\Gamma_{B}}\frac{d\Gamma}{dq^2}\,,
\label{dgamma}\\[2ex]
A_{\rm FB} &=& 
\frac{1}{d\Gamma/dq^2} \left[ \int\limits_0^1 - \int\limits_{-1}^0 \right]
d\!\cos\theta\, \frac{d^2\Gamma}{dq^2 d\!\cos\theta} 
= - \frac34 \frac{J_{6s}}{d\Gamma/dq^2}
= -\frac34\beta_\ell \frac{ {\cal H}_P^{12}} { {\cal H}_{\rm tot} }\,,
\label{AFB}\\[2ex]
F_L &=& -\frac{J_{2c}}{d\Gamma/dq^2} = 
\frac12 \beta_\ell^2 \frac{ {\cal H}_L^{11} +  {\cal H}_L^{22}}{{\cal H}_{\rm tot} }.
}

%--- observables from 1303.5794 -------------------

Then, inspired by Ref.~\cite{Descotes-Genon:2013vna}, we 
introduce the various observables 
integrated over the relevant kinematic range
\bea
\av{P_1}_{\rm bin} &=& \frac12 \frac{\int_{{\rm bin}} dq^2 J_3}
                                {\int_{{\rm bin}} dq^2 J_{2s}}
= -2\, \frac{\int_{{\rm bin}} dq^2\,\beta_\ell^2\,f(q^2) 
                                [{\cal H}_T^{11}+ {\cal H}_T^{22}]}
          {\int_{{\rm bin}} dq^2\,\beta_\ell^2\, f(q^2)
                                [{\cal H}_U^{11}+ {\cal H}_U^{22}] },
\nn[2ex]
\av{P_2}_{\rm bin} &=& \frac18 \frac{\int_{{\rm bin}} dq^2 J_{6s}}    
                                 {\int_{{\rm bin}} dq^2 J_{2s}}
 = -\,\frac{\int_{{\rm bin}} dq^2\,\beta_\ell\, f(q^2) {\cal H}_P^{12}}
          {\int_{{\rm bin}} dq^2\,\beta_\ell^2\, 
                                [{\cal H}_U^{11}+ {\cal H}_U^{22}] },
\nn[2ex]
\nn
\av{P_3}_{\rm bin} &=& -\frac14 \frac{\int_{{\rm bin}} dq^2 J_9}
                                  {\int_{{\rm bin}} dq^2 J_{2s}}
= -\,\frac{\int_{{\rm bin}} dq^2\,\beta_\ell^2\,f(q^2) 
                                  [{\cal H}_{IT}^{11}+ {\cal H}_{IT}^{22}]}
          {\int_{{\rm bin}} dq^2\,\beta_\ell^2\, f(q^2)
                                [{\cal H}_U^{11}+ {\cal H}_U^{22}] },
\nn[2ex]
\av{P'_4}_{\rm bin} &=& \frac1{{\cal N}_\bin} \int_{{\rm bin}} dq^2 J_4
= 2\,\frac{\int_{{\rm bin}} dq^2\,\beta_\ell^2\,f(q^2) 
                         [{\cal H}_I^{11}+ {\cal H}_I^{22}]}{N_{\rm bin}},
\nn[2ex]
\av{P'_5}_{\rm bin} &=& \frac1{2{\cal N}_\bin} \int_{{\rm bin}} dq^2 J_5
= -2\,\frac{\int_{{\rm bin}} dq^2\,\beta_\ell\,f(q^2) 
                           [{\cal H}_A^{12}+ {\cal H}_A^{21}]}{N_{\rm bin}},
\nn[2ex]
\av{P'_6}_{\rm bin} &=& \frac{-1}{2{\cal N}_\bin} \int_{{\rm bin}} dq^2 J_7
= -2\,\frac{\int_{{\rm bin}} dq^2\,\beta_\ell\, f(q^2)
                         [{\cal H}_{II}^{12}+ {\cal H}_{II}^{21}]}{N_{\rm bin}},
\nn[2ex]
\av{P'_8}_{\rm bin} &=& \frac{-1}{ {\cal N}_\bin} \int_{{\rm bin}} dq^2 J_8
= +2\,\frac{\int_{{\rm bin}} dq^2\,\beta_\ell^2\, f(q^2)
                         [{\cal H}_{IA}^{11}+ {\cal H}_{IA}^{22}]}{N_{\rm bin}},
\label{binP1-8}
\ena
where the normalization ${\cal N}_\bin$ is defined as
\bea
{\cal N}_\bin &=& {\textstyle \sqrt{-\int_\bin dq^2 [J_{2s}]\cdot 
                                            \int_{{\rm bin}} dq^2 [J_{2c}]}}
\nn[2ex]
&=& {\textstyle \sqrt{\int_\bin dq^2\,\beta_\ell^2\,f(q^2) 
                                 [{\cal H}_U^{11}+{\cal H}_U^{22}]\cdot 
                    \int_{{\rm bin}}\,dq^2\,\beta_\ell^2\, f(q^2)
                                 [{\cal H}_L^{11}+{\cal H}_L^{22}]}}.
\label{Nbin}
\ena
We also use the phase space factor 
$f(q^2)=|{\bf p_2}|\,q^2\,\beta_\ell$
in the numerator and denominator when calculating the $q^2$-averages. 

%-------------------- insertion -----------------------------------

One should notice that the observables $P_i$ are linked to the full
four-fold distribution Eq.~(\ref{eq:distr4}), because the corresponding
helicity amplitudes (or equivalent $J_i$ coefficients) do not appear in the
single or double differential distributions 
Eqs.~(\ref{eq:distr1}) and (\ref{eq:distr2}).

%\clearpage
\section{Numerical results}
\label{sec:res}

In this section we present the numerical results obtained for
various physical observables. The values of the lepton and meson masses
and the B-meson lifetime  are taken from Ref.~\cite{Agashe:2014kda}.
%
%----------------- insertion ------------------
%
The SM Wilson coefficients are taken from Ref.~\cite{Descotes-Genon:2013vna}.
They were computed at the matching scale $\mu_0=2 M_W$ and run down to 
the hadronic scale $\mu_b= 4.8$~GeV.
The evolution of couplings and current quark masses proceeds analogously. 
%--------- end insertion -----------------
The values of the model independent input parameters
and the Wilson coefficients are listed in Table~\ref{tab:input}.
\begin{table}[htbp]
\caption{Values of the input parameters and the corresponding
values of the Wilson coefficients used in the numerical calculations.}  
\centering
\def\arraystretch{1.5}
 \begin{tabular}{ccccccccc}
\hline\hline
  $m_W$ &  $\sin^2\theta_W $ &  $\alpha(M_Z)$ & 
$\bar m_c$ &  $\bar m_b$  &  $\bar m_t$ & $\lambda_t=|V^\dagger_{ts} V_{tb}|$  & &\\
\hline
 $80.41$~GeV & $0.2313$ & $1/128.94$ & $1.27$~GeV & $4.68$~GeV & $173.3$~GeV &  
 0.041 & & \\
\hline\hline
 $C_1$ &  $C_2$ &  $C_3$ &  $C_4$ & $C_5$ &  $C_6$ &  $C^{\rm eff}_7$ &   
$C_9$ &  $C_{10}$ \\
\hline
 $-0.2632$ & $1.0111$ &  $-0.0055$ & $-0.0806$ & 0.0004 & 0.0009 & $-0.2923$ &
  4.0749 & $-4.3085$ \\
\hline\hline
 \end{tabular}
\label{tab:input}
\end{table}

In Table~\ref{tab:branching} we give the numerical values for
the total branching ratios and compare them with available
experimental data.
\begin{table}[ht]
\begin{center}
\caption{Total branching fractions.}
\begin{tabular}{c|cccc}
\hline
Mode & Our & \multicolumn{2}{c}{Others} &
Expt.~\cite{Agashe:2014kda,Aaij:2014pli,Pescatore:2014ksa} \\
\hline
$B\to K^\ast\mu^+\mu^- $ & $12.7\times 10^{-7}$ & $(11.9 \pm 3.9) \times
10^{-7}$ &\cite{Ali:2002jg} &
$(9.24\pm 0.93 ({\rm stat}) \pm 0.67 ({\rm sys}))\times 10^{-7}$ \\
& & $19 \times 10^{-7}$ &\cite{Ali:1999mm} & \\
& & $11.5 \times 10^{-7}$ &\cite{Melikhov:1997wp}& \\
& & $14 \times 10^{-7}$ &\cite{Geng:1996az}& \\
$B\to K^\ast\tau^+\tau^- $ & $1.35\times 10^{-7}$ & $1.9 \times 10^{-7} $
&\cite{Ali:1999mm}& $-$ \\
& & $1.0 \times 10^{-7}$ &\cite{Melikhov:1997wp}& \\
& & $2.2 \times 10^{-7}$ &\cite{Geng:1996az}& \\
$B\to K^\ast\gamma $ & $3.74\times 10^{-5}$ & $11.4 \times 10^{-5}$ &
\cite{Ivanov:1998ms} &
$(4.21\pm 0.18)\times 10^{-5}$ \\
& & $4.2 \times 10^{-5}$ &\cite{Melikhov:1997wp}& \\
$B\to K^\ast\nu\bar\nu $ & $1.36\times 10^{-5}$ & $1.5 \times 10^{-5}$ &
\cite{Melikhov:1997wp} & $-$ \\
\hline
$B\to K\mu^+\mu^- $ & $7.18\times 10^{-7}$ & $5.7 \times 10^{-7}$ &
\cite{Ali:1999mm} &
$(4.29\pm 0.07 ({\rm stat}) \pm 0.21 ({\rm sys}))\times 10^{-7}$ \\
& & $(3.5 \pm 1.2) \times 10^{-7}$ & \cite{Ali:2002jg} & \\
& & $4.4 \times 10^{-7}$ & \cite{Melikhov:1997wp} & \\
& & $5 \times 10^{-7}$ & \cite{Geng:1996az} & \\
$B\to K\tau^+\tau^- $ & $3.0\times 10^{-7}$ & $1.3 \times 10^{-7}$ &
\cite{Ali:1999mm} & $-$ \\
& & $1.0 \times 10^{-7}$ & \cite{Melikhov:1997wp} & \\
& & $1.3 \times 10^{-7}$ & \cite{Geng:1996az} & \\
$B\to K\nu\bar\nu $ & $0.60\times 10^{-5}$ & $0.56 \times 10^{-5}$ &
\cite{Melikhov:1997wp} & $-$ \\
\hline
\end{tabular}
\label{tab:branching}
\end{center}
\end{table}

\begin{figure*}[htbp]
\centering
\begin{tabular}{lr}
\includegraphics[scale=0.6]{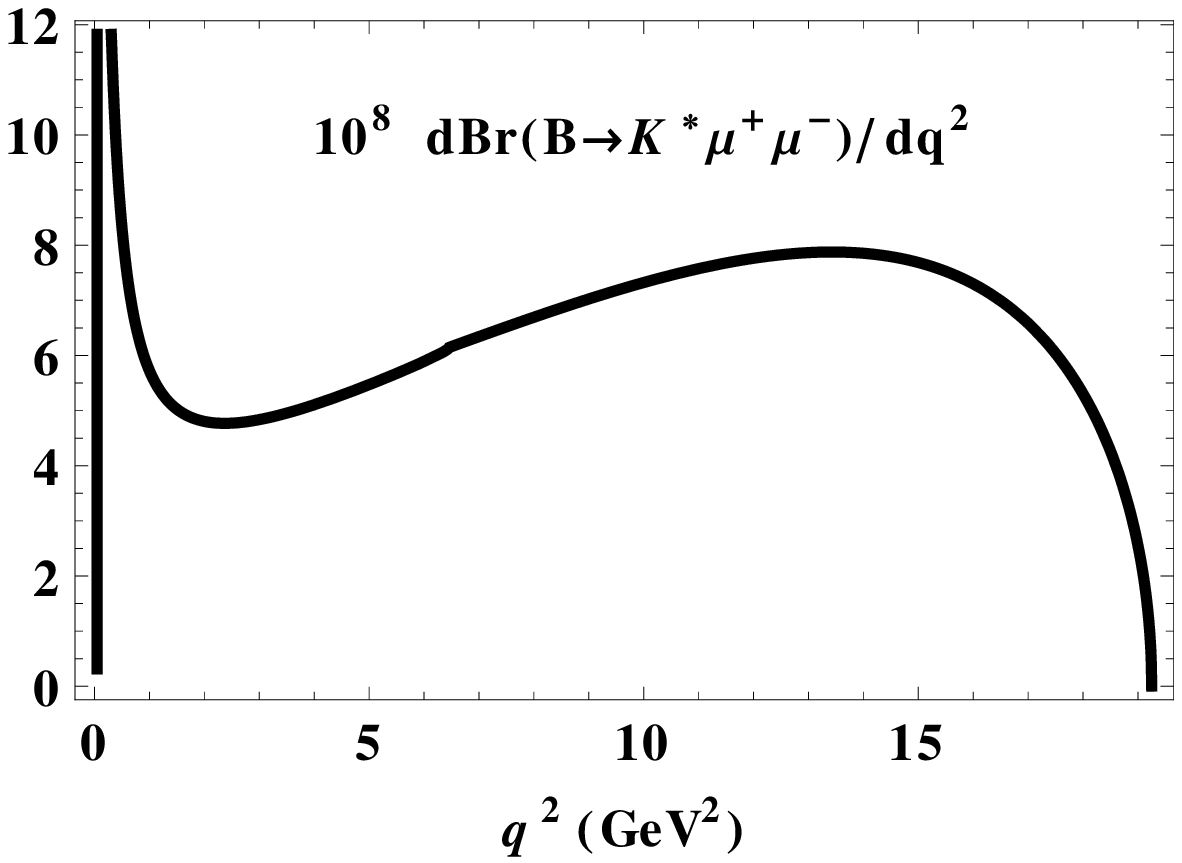} &
\includegraphics[scale=0.6]{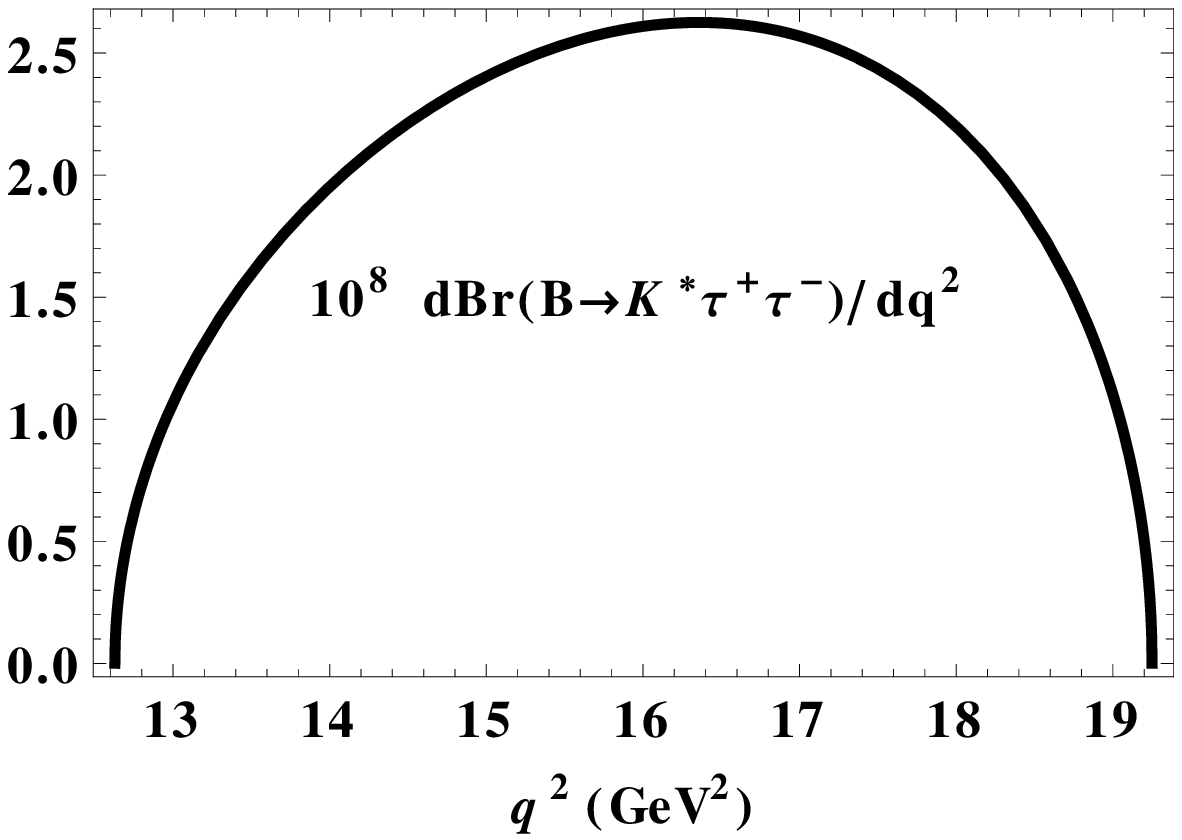}\\
\end{tabular}
\caption{Differential decay widths of the decays $B \to K^\ast \ell^{+} \ell^{-}$.}
\label{fig:difwidths}
\end{figure*}

\begin{figure*}[htbp]
\centering
\begin{tabular}{lr}
\includegraphics[scale=0.6]{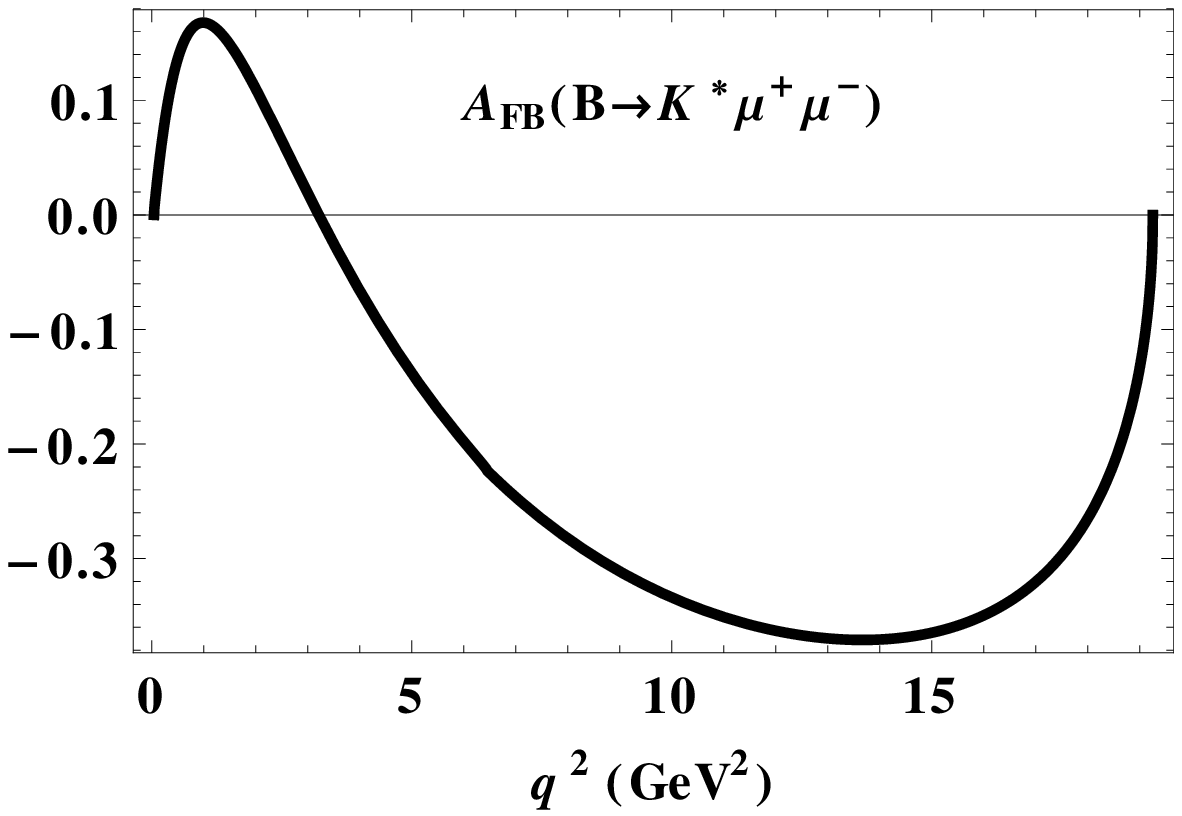} &
\includegraphics[scale=0.6]{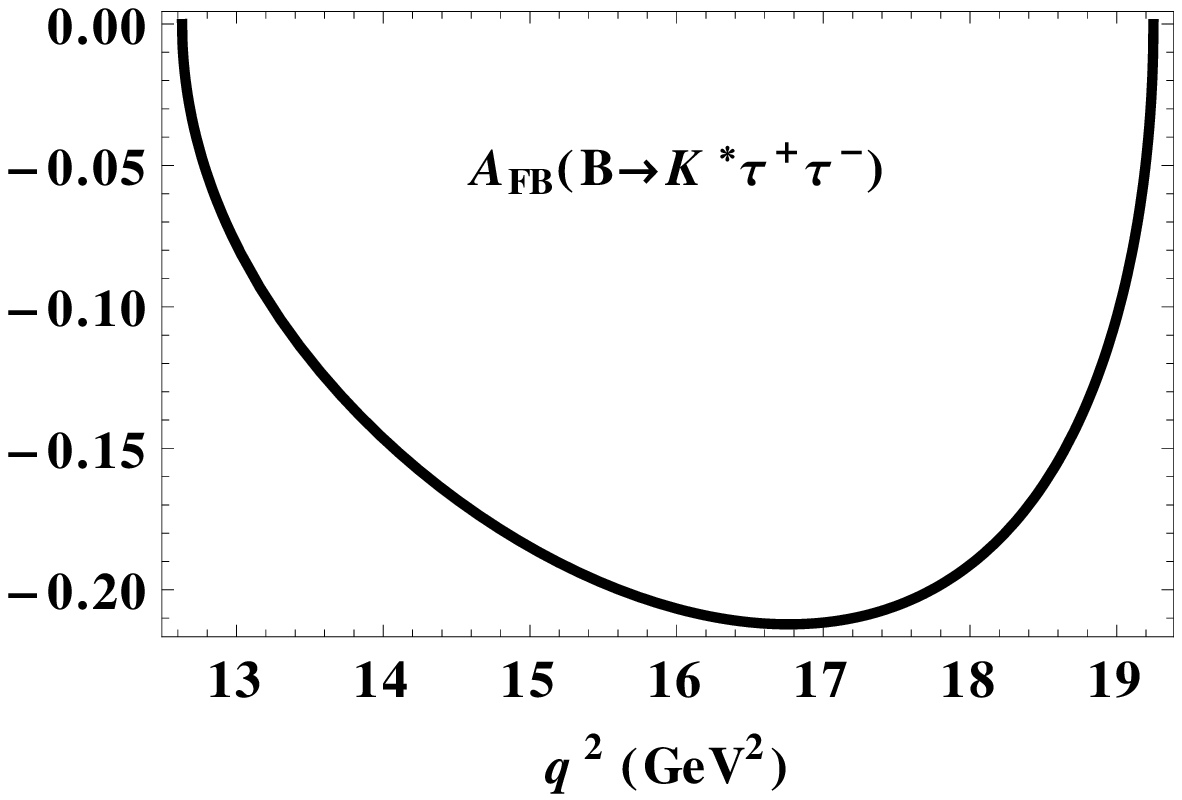}\\
\end{tabular}
\caption{Forward-backward asymmetry for the decays $B \to K^\ast \ell^{+} \ell^{-}$.}
\label{fig:AFB}
\end{figure*}

\begin{figure*}[htbp]
\centering
\begin{tabular}{lr}
\includegraphics[scale=0.6]{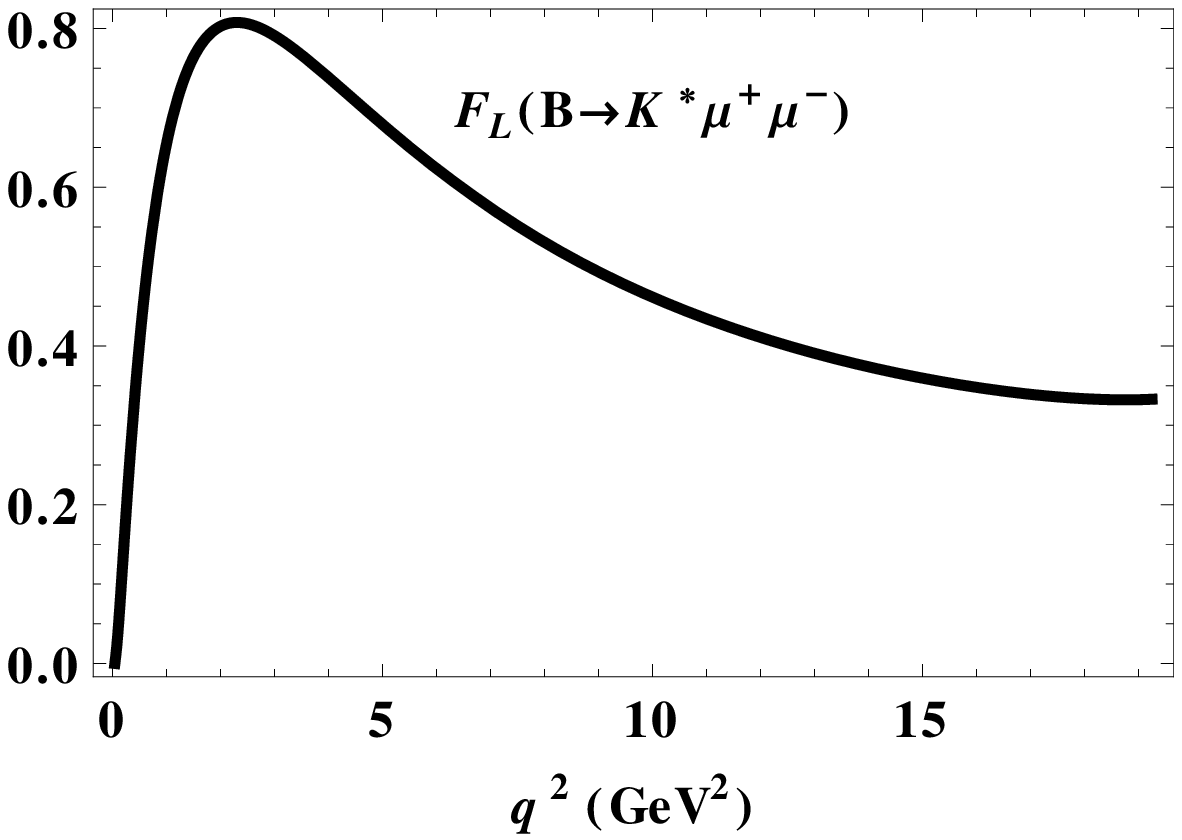} &
\includegraphics[scale=0.6]{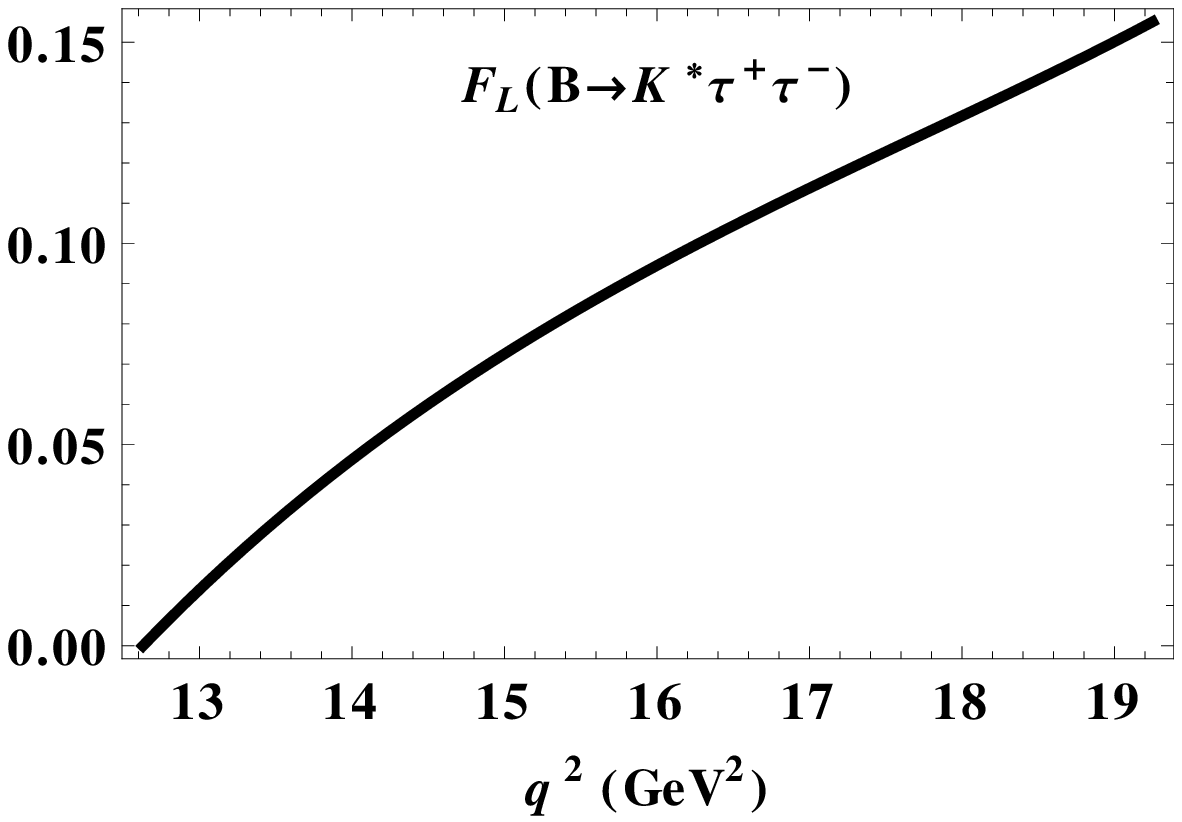}\\
\end{tabular}
\caption{Longitudinal polarization for the decays $B \to K^\ast \ell^{+} \ell^{-}$.}
\label{fig:FL}
\end{figure*}

\begin{figure*}[htbp]
\centering
\begin{tabular}{lr}
\includegraphics[scale=0.6]{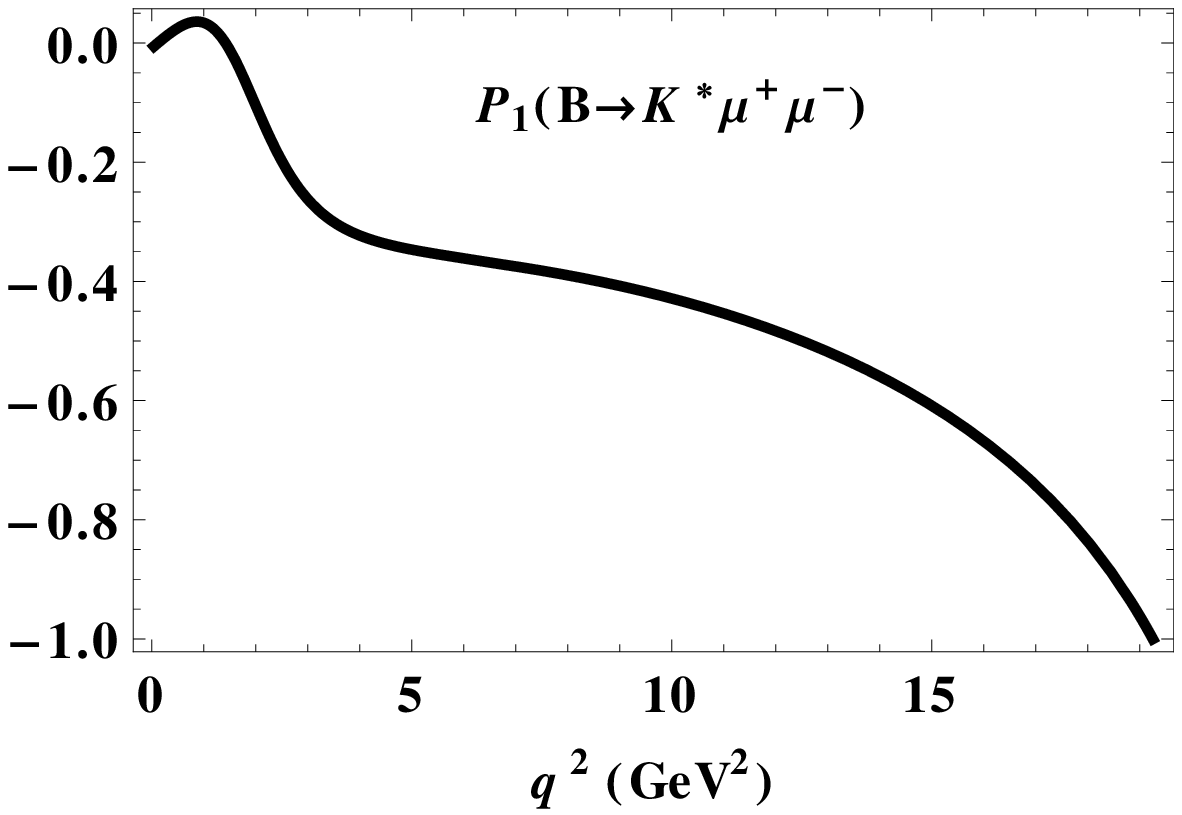} &
\includegraphics[scale=0.6]{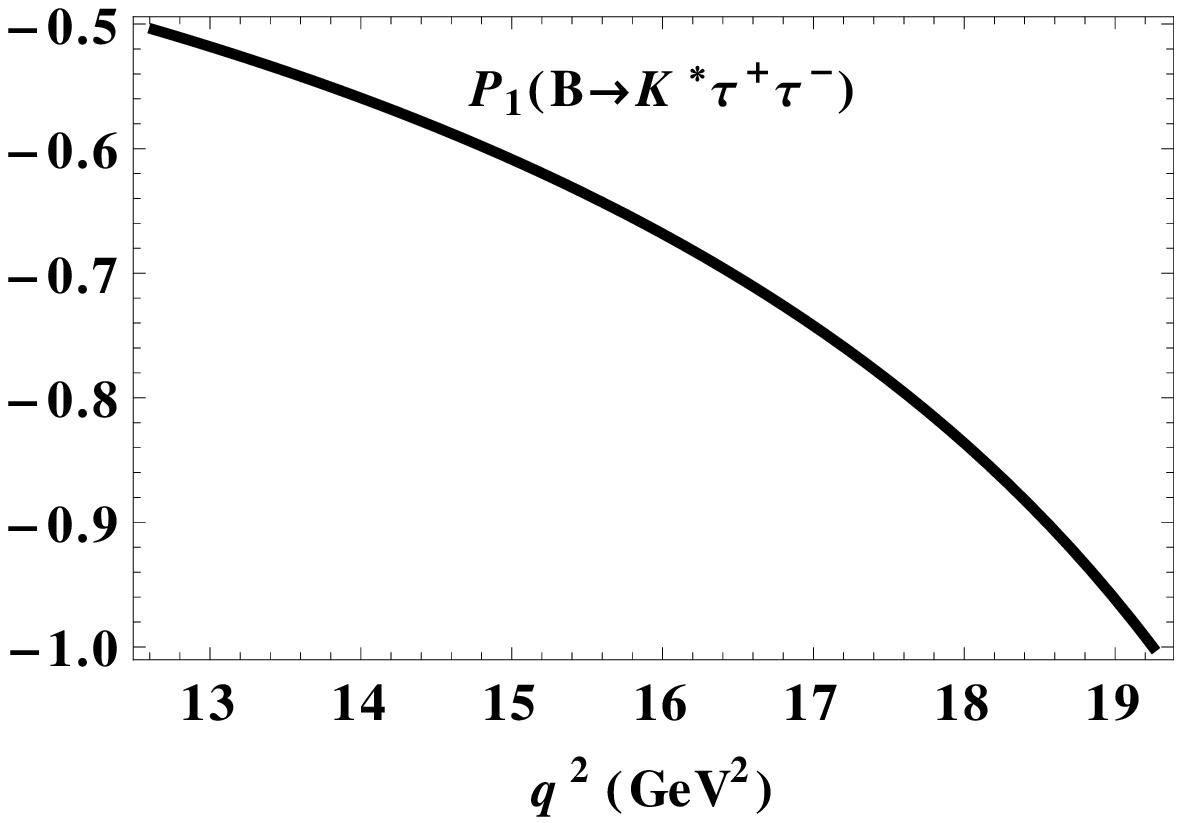}\\
\includegraphics[scale=0.6]{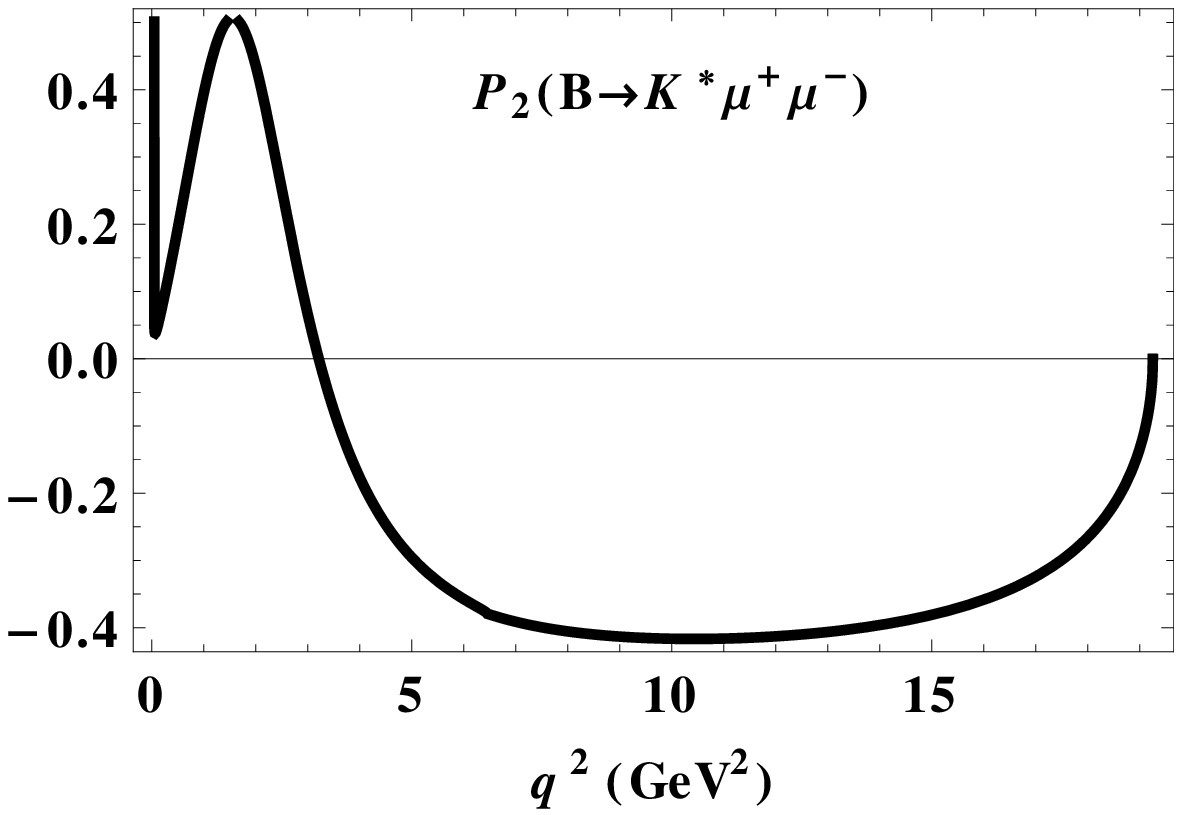} &
\includegraphics[scale=0.6]{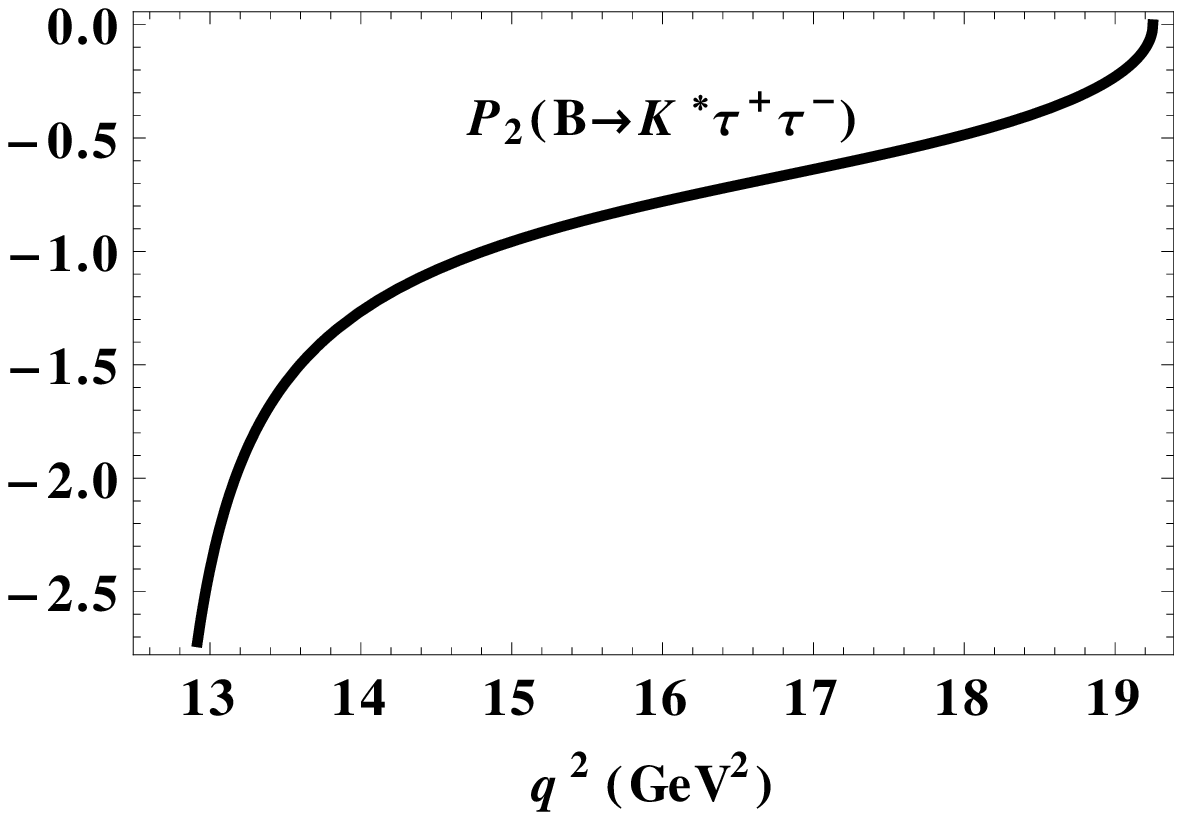}\\
\includegraphics[scale=0.6]{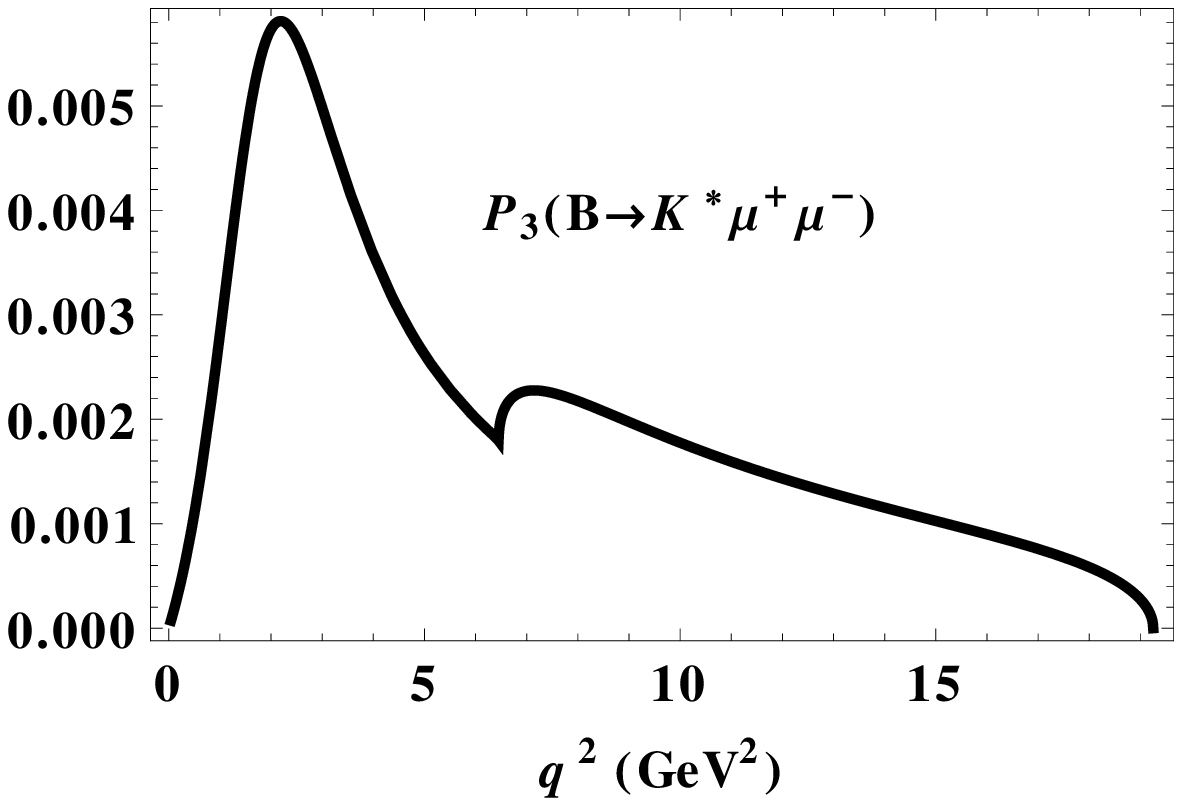} &
\includegraphics[scale=0.6]{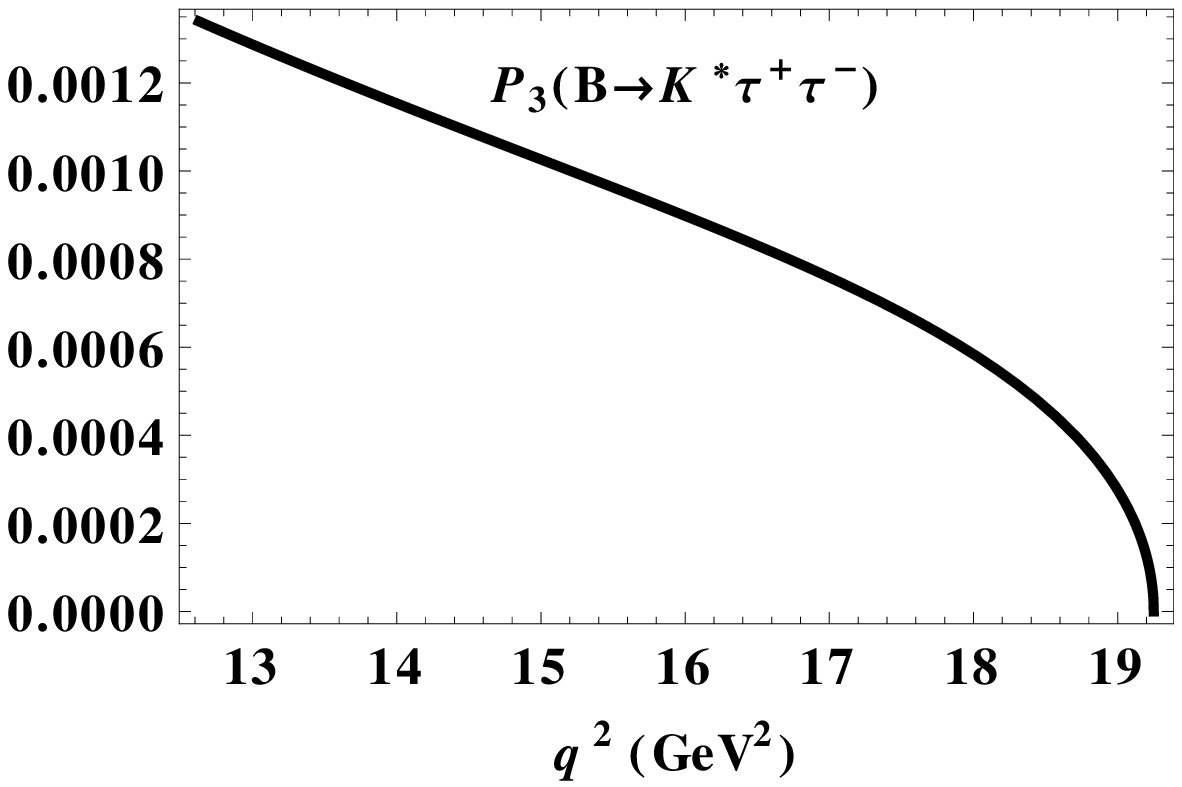}\\
\end{tabular}
\caption{Clean observables $P_{1,2,3}$ for the decays $B \to K^\ast \ell^{+} \ell^{-}$.}
\label{fig:P123}
\end{figure*}

\begin{figure*}[htbp]
\centering
\begin{tabular}{lr}
\includegraphics[scale=0.6]{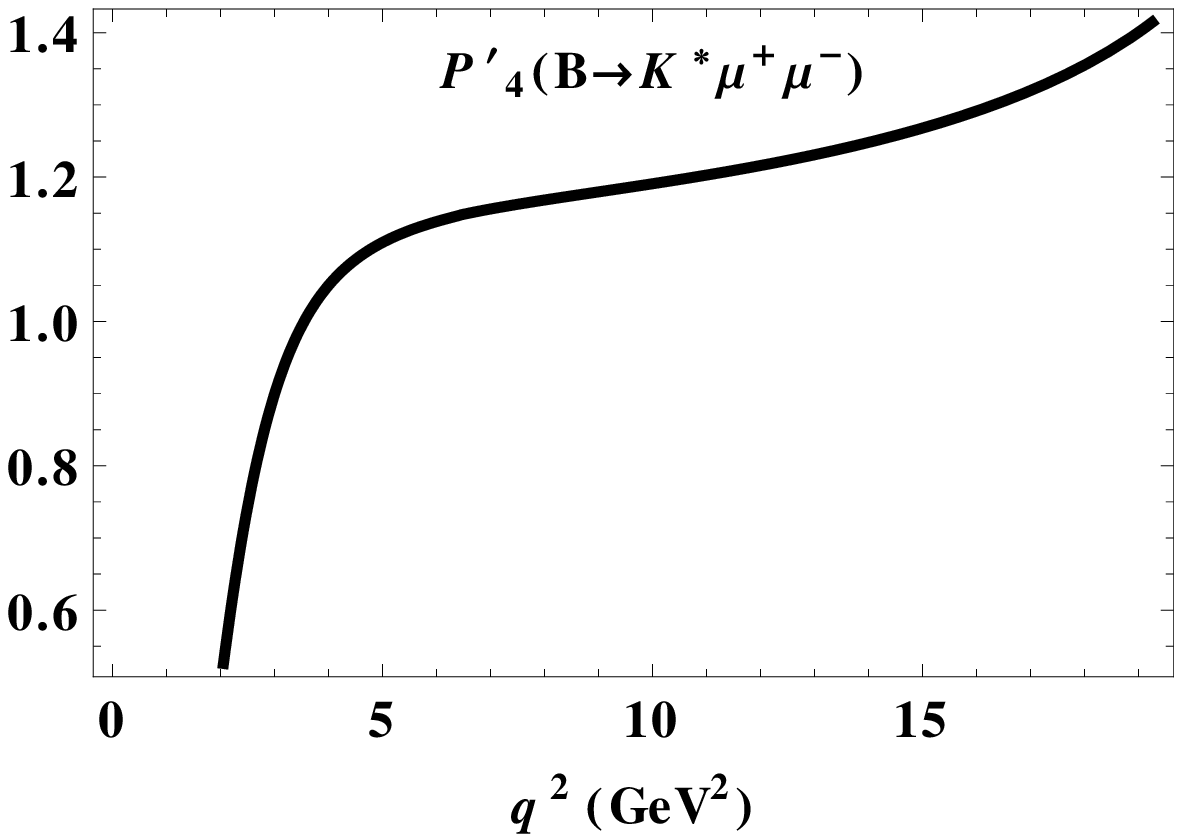} &
\includegraphics[scale=0.6]{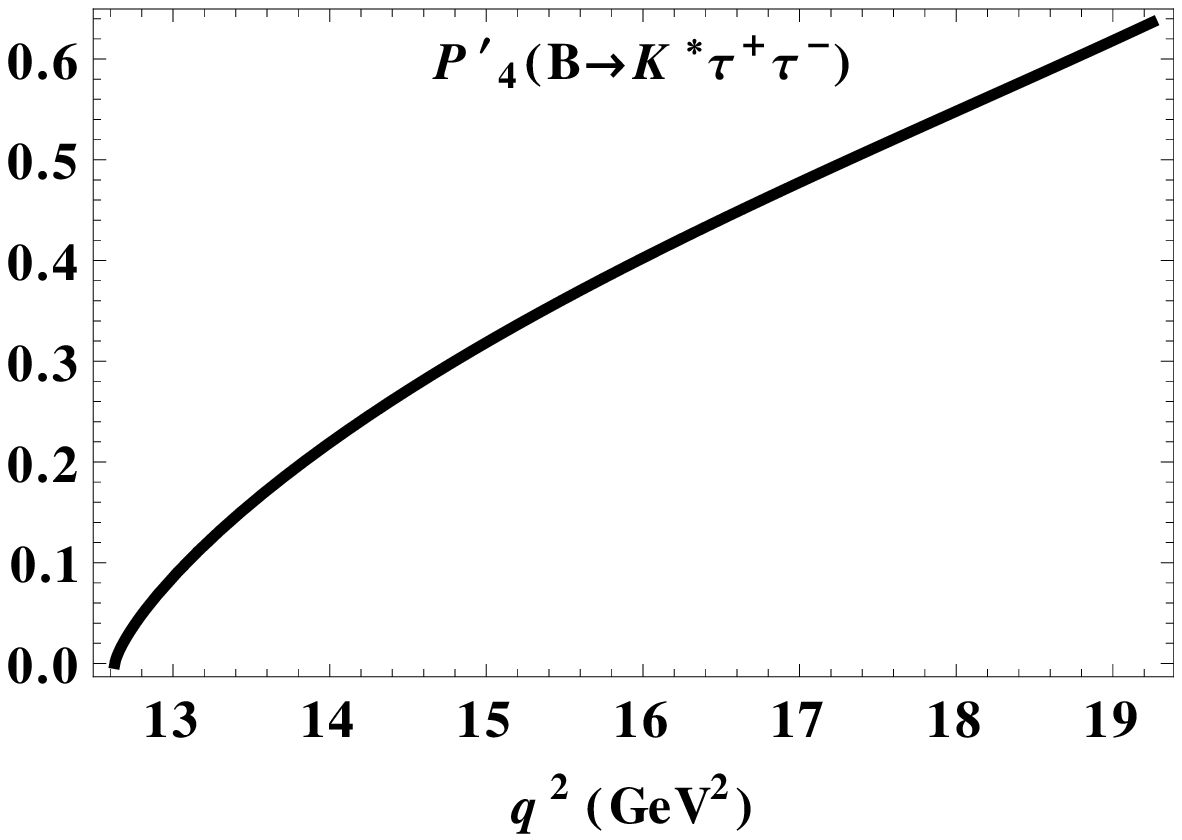}\\
\includegraphics[scale=0.6]{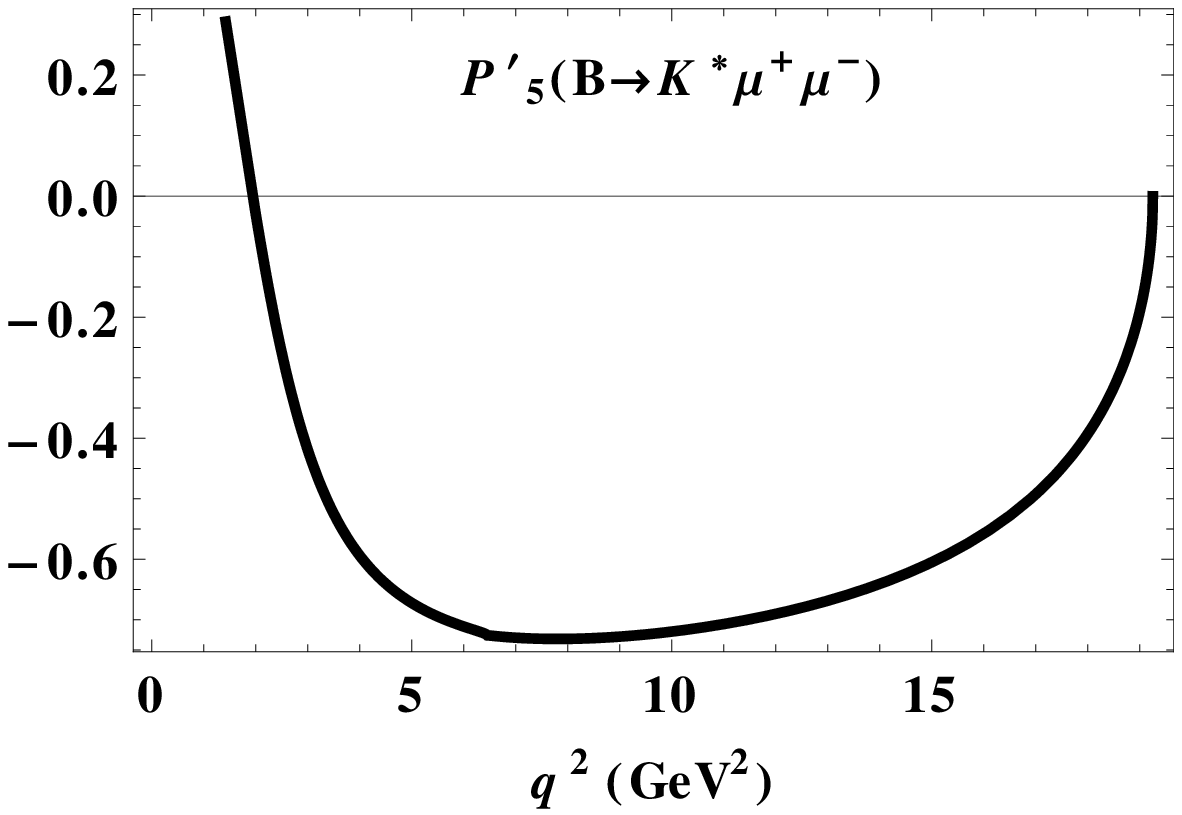} &
\includegraphics[scale=0.6]{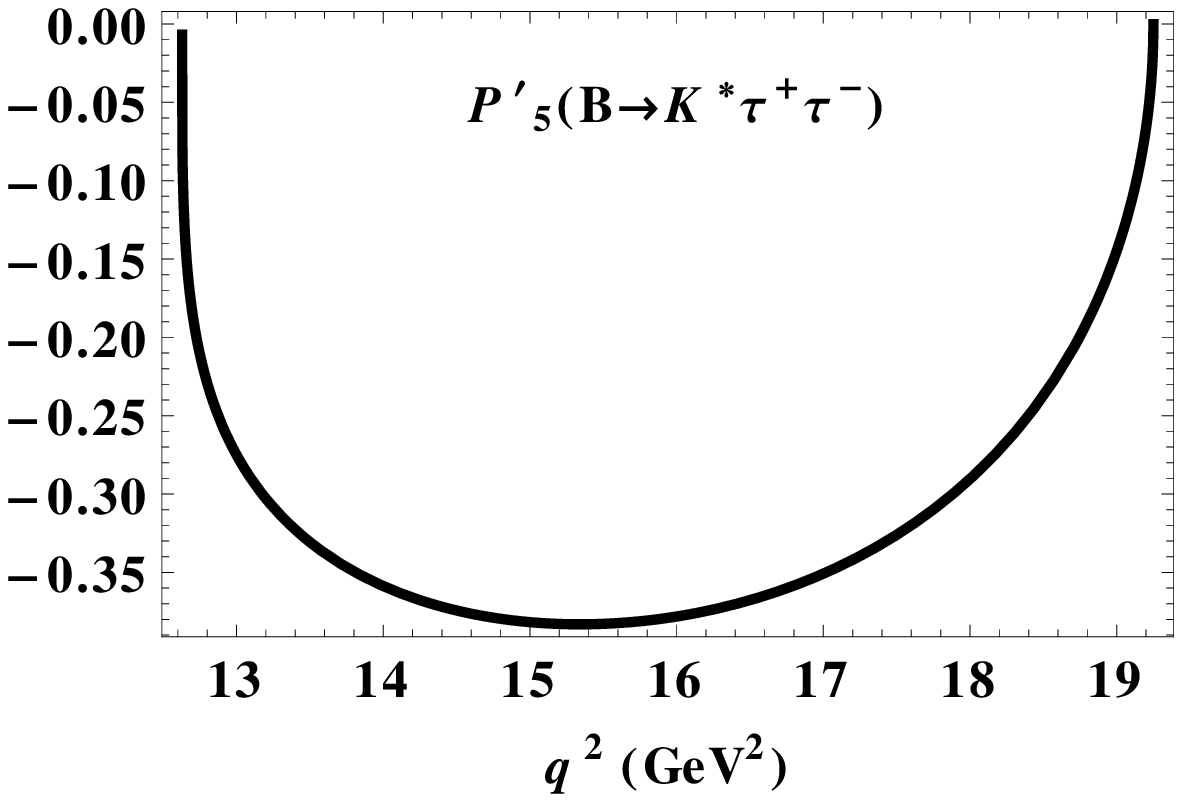}\\
\includegraphics[scale=0.6]{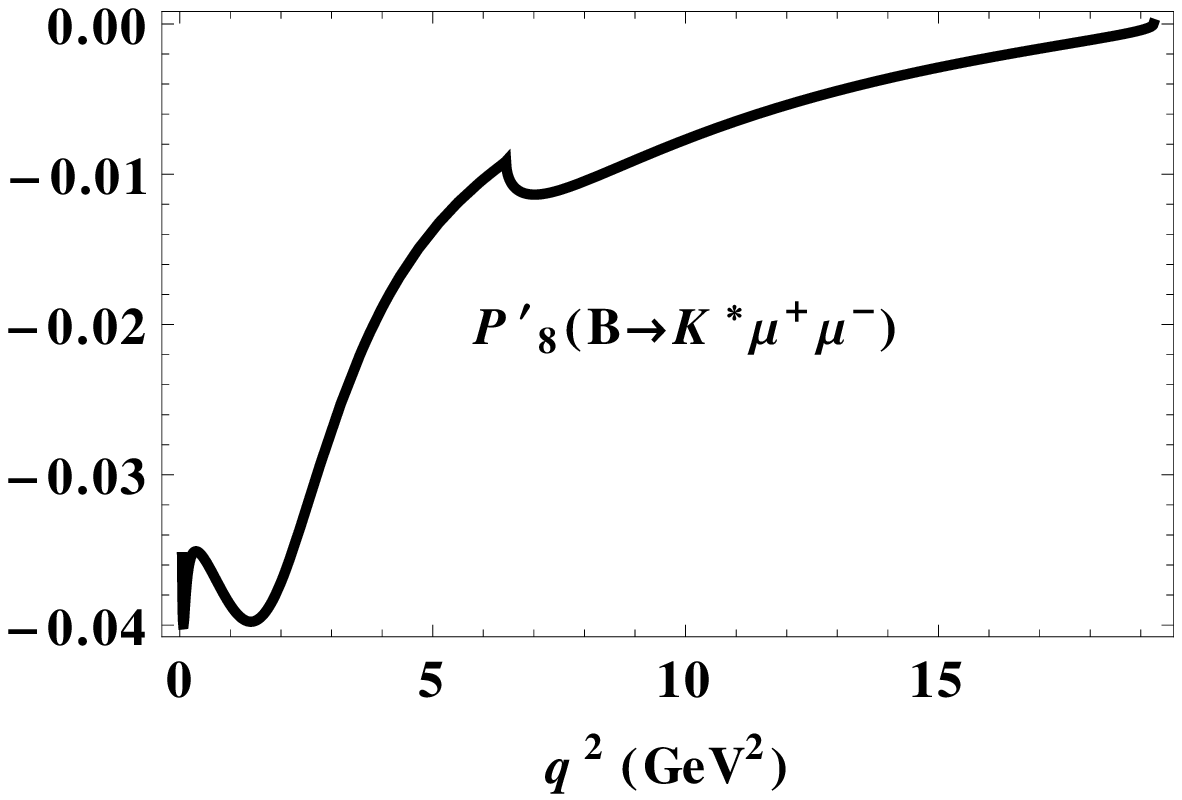} &
\includegraphics[scale=0.6]{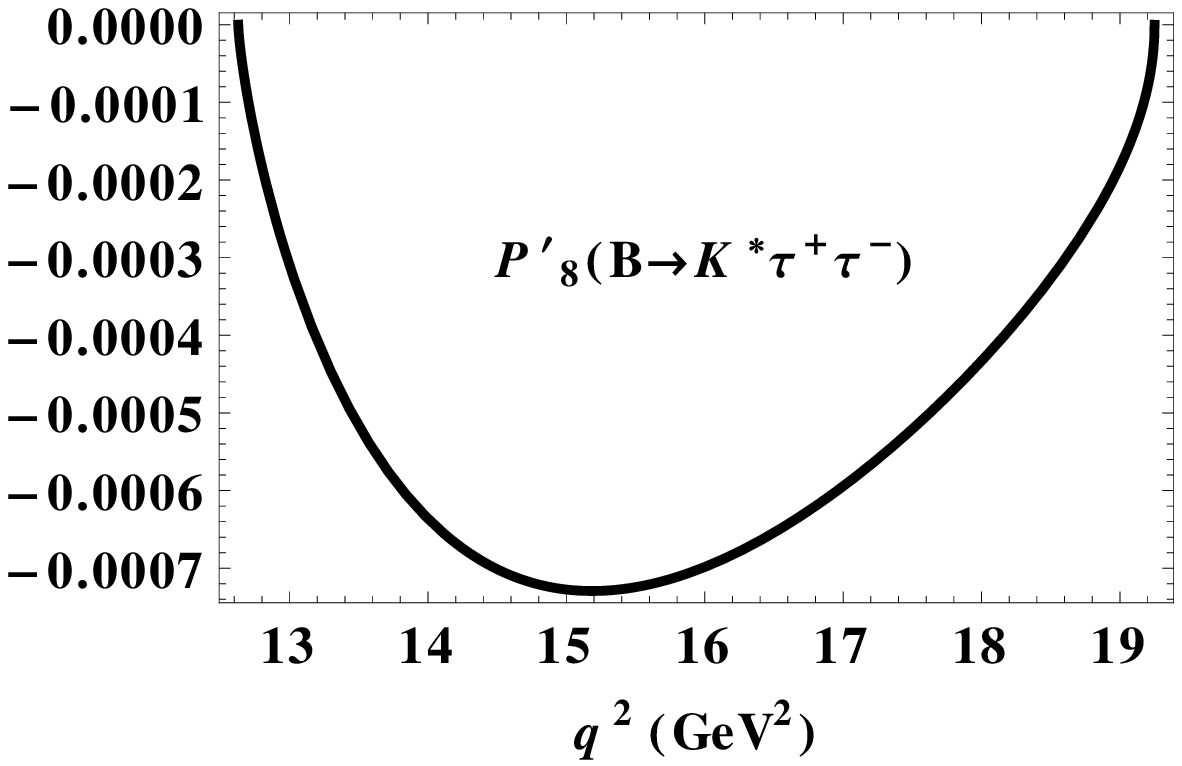}\\
\end{tabular}
\caption{Clean observables $P'_{4,5,8}$ for the decays $B \to K^\ast \ell^{+} \ell^{-}$.}
\label{fig:P458}
\end{figure*}

In Figs.~\ref{fig:difwidths}, \ref{fig:AFB} and \ref{fig:FL}
we display our results for the differential decay widths,
the forward-backward asymmetry and the longitudinal polarization
for the decay  $B\to K^\ast\ell^+\ell^-$ in the full kinematical region.
We also plot unintegrated clean observables $P_i$ in Figs.~\ref{fig:P123}
and \ref{fig:P458}. The $q^2$-averages of all polarization observables 
are given in Table~\ref{tab:obs-numerics}.  
\begin{table}[ht] 
\begin{center}
\def\arraystretch{1.2}
\begin{tabular}{c|cccccccc}
\hline
\multicolumn{8}{c}{ $B\to K^\ast\ell^+\ell^-$ }\\
\hline
& $<A_{FB}>$       
& $<F_L>$ 
& $<P_1>$   
& $<P_2>$ 
& $<P_3>$ 
& $<P_4'>$ 
& $<P_5'>$ 
& $<P_8'>$ 
\\
\hline
 $\mu$   
& $ -0.23 $  
& $  0.47 $  
& $ -0.48 $   
& $ -0.31 $ 
& $  0.0015 $   
& $  1.01 $   
& $ -0.49 $   
& $ -0.010 $   
\\
 $\tau$     
& $ -0.18 $   
& $  0.092 $  
& $ -0.74 $    
& $ -0.68 $  
&  $ 0.00076 $   
&  $ 1.32 $   
&  $-1.07 $   
&  $-0.0018 $   
\\
\hline
\end{tabular}
\vspace{0.5cm}
\caption{$q^{2}$--averages of polarization observables over the whole 
allowed kinematic region. }
\label{tab:obs-numerics}
\end{center}
\end{table}

Finally, we present our results for the binned observables
in Tables~\ref{tab:bin-1} and ~\ref{tab:bin-2}. The binning is chosen
to match the current experimental data  \cite{0904.0770,LHCbinned,1108.0695} 
and the results from other theoretical approaches \cite{Descotes-Genon:2013vna}.

The $q^2$-integrated predictions and measurements ($1-6 \, \mathrm{GeV^2}$) for
branching fraction and $A_{FB}$ and $F_L$ observables stand
\begin{center}
\def\arraystretch{1.2}
\begin{tabular}{c|c c c c}
 & \quad Belle \cite{0904.0770} \quad & \quad LHCb \cite{LHCbinned}  \quad & 
 \quad CDF \cite{1108.0695} \quad & 
 \quad CQM \tabularnewline
\hline 
$\mathcal{B}\times10^{7}$  \quad &  \quad $1.49_{-0.40}^{+0.45}\pm0.12$  \quad & 
 \quad $0.42\pm0.06\pm0.03$  \quad &  \quad  - & $2.58$
\tabularnewline
$A_{FB}$  \quad &  \quad $0.26_{-0.30}^{+0.27}\pm0.07$  \quad &  
\quad $-0.06_{-0.14}^{+0.13}\pm0.04$  \quad & \quad  
$0.29_{-0.23}^{+0.20}\pm0.07$ & $-0.02$
\tabularnewline
$F_{L}$  \quad & $0.67_{-0.23}^{+0.23}\pm0.05$  \quad &  
\quad $0.55\pm0.10\pm0.03$  \quad &  \quad  $0.69_{-0.21}^{+0.19}\pm0.08$  \quad &
 \quad  $0.75$
\tabularnewline
\end{tabular}
\par\end{center}

Considering the outcoming numbers, it is difficult to make a clear statement 
about the level of agreement. Clearly, the branching fraction prediction 
is above both measured values. On the other hand one has to note 
the discrepancy which exists between the experimental values themselves 
and thus a question mark remains on this issue. Concerning the two 
other observables, $A_{FB}$ and $F_L$, the model predictions are in agreement 
with the experiments (slightly overshooting the $F_L$ value for 
the LHCb experiment), the measurement errors being however quite important. 
Basically the same commentary can be made about the results in other bins. 
It is probably necessary to wait until the measurement errors shrink so that 
the experiments become more constraining.
\begin{table*}[htb]
\caption{
Binned observables ${\cal B}$, $F_L$ and $A_{FB}$ for 
$B \to K^* \mu^+ \mu^-$ (CQM - covariant quark model).
}
\label{tab:bin-1}
\begin{center}
\begin{tabular}{c|ccccc}
\hline\hline
Bin ($GeV^{2}$) & \cite{0904.0770}  & \cite{LHCbinned} & 
\cite{1108.0695} & \cite{Descotes-Genon:2013vna} & CQM  \\
\hline
& & & ${\mathcal B}$(10$^{-7}$) & & \\
\hline
1.00--2.00      & 
$-$ & 
$-$ & 
$-$ &  
$0.437_{-0.148 - 0.023}^{+0.345 + 0.026}$ & 
$0.51$ \\
0.00--2.00      & 
1.46$^{+0.40}_{-0.35}\pm$0.11~ & 
$0.61 \pm 0.12\pm 0.06$ & 
- &  
$1.446_{-0.561 - 0.054}^{+1.537 + 0.057}$ & 
$1.40$ \\
2.00--4.30      & 
0.86$^{+0.31}_{-0.27}\pm$0.07~ & 
$0.34 \pm 0.09\pm 0.02$ & 
- & 
$0.904_{-0.314 - 0.055}^{+0.664 + 0.061}$ & 
1.13 \\
4.30--8.68    & 
1.37$^{+0.47}_{-0.42}\pm$0.39~ & 
$0.69 \pm 0.08\pm 0.05$ & 
- & 
$2.674_{-0.973 - 0.145}^{+2.326 + 0.156}$ & 
2.67 \\
10.09--12.89 & 
2.24$^{+0.44}_{-0.40}\pm$0.19~ & 
$0.55 \pm 0.09\pm 0.07$ & 
- & 
$2.344_{-1.100 - 0.063}^{+2.814 + 0.069}$ & 
2.14 \\
14.18--16.00    & 
1.05$^{+0.29}_{-0.26}\pm$0.08~ & 
$0.63 \pm 0.11\pm 0.05$ & 
- & $1.290_{-0.815 - 0.013}^{+2.122 + 0.013}$ & 
1.39 \\
$>$16.00       & 
2.04$^{+0.27}_{-0.24}\pm$0.16~ & 
$0.50 \pm 0.08\pm 0.05$ & 
-  & 
$1.450_{-0.922 - 0.015}^{+2.333 + 0.015}$ &  
1.71 \\
1.00--6.00      & 
1.49$^{+0.45}_{-0.40}\pm$0.12~ & $0.42 \pm 0.06\pm 0.03$ &  
- & 
$2.155_{-0.742 - 0.123}^{+1.646 + 0.138}$ & 
2.58 \\
\hline\hline
& & & $A_{FB}$ & & \\
\hline
1.00--2.00      & 
$-$ & 
$-$ & 
$-$ & 
$-0.212_{-0.144 - 0.015}^{+0.11 + 0.014}$ & 
$-0.15$ \\
0.00--2.00      & 
0.47$^{+0.26}_{-0.32}\pm$0.03 &  
$-0.15\pm 0.20\pm 0.06$ & 
$-0.35^{+0.26}_{-0.23}\pm 0.10$ & 
$-0.136_{-0.045 - 0.016}^{+0.048 + 0.016}$ & 
$-0.12$ \\
2.00--4.30      & 
$0.37^{+0.25}_{-0.24}\pm 0.10$  & 
$\phantom-0.05~\,^{+\,0.16}_{-\,0.20}\,\pm 0.04$ & 
$0.29^{+0.32}_{-0.35}\pm 0.15$ & 
$-0.081_{-0.068 - 0.009}^{+0.054 + 0.008}$  & 
$-0.0059$ \\
4.30--8.68    & 
0.45$^{+0.15}_{-0.21}\pm$0.15 & 
$\phantom-0.27~\,^{+\,0.06}_{-\,0.08}\,\pm 0.02$ & 
$0.01^{+0.20}_{-0.20}\pm 0.09$ & 
$0.220_{-0.112 - 0.016}^{+0.138 + 0.014}$ & 
$0.22$ \\
10.09--12.89 & 
0.43$^{+0.18}_{-0.20}\pm$0.03 & 
$\phantom-0.27~\,^{+\,0.11}_{-\,0.13}\,\pm 0.02$ &  
$0.38^{+0.16}_{-0.19}\pm 0.09$ & 
$0.371_{-0.164 - 0.011}^{+0.150 + 0.010}$ & 
$0.36$ \\
14.18--16.00    & 
0.70$^{+0.16}_{-0.22}\pm$0.10 & 
$\phantom-0.47~\,^{+\,0.06}_{-\,0.08}\,\pm 0.03$ & 
 $0.44^{+0.18}_{-0.21}\pm 0.10$ &  
$0.404_{-0.191 - 0.005}^{+0.199 + 0.005}$ & 
$0.36$ \\
$>$16.00       &  
0.66$^{+0.11}_{-0.16}\pm$0.04  & 
$\phantom-0.16~\,^{+\,0.11}_{-\,0.13}\,\pm 0.06$ & 
$0.65^{+0.17}_{-0.18}\pm 0.16$  & 
$0.360_{-0.172 - 0.005}^{+0.205 + 0.004}$ &  
$0.29$ \\
1.00--6.00      &  
0.26$^{+0.27}_{-0.30}\pm$0.07 & 
$-0.06~\,^{+\,0.13}_{-\,0.14}\,\pm 0.04$ & 
 $0.29^{+0.20}_{-0.23}\pm 0.07$  & 
$-0.035_{-0.033 - 0.009}^{+0.036 + 0.008}$ & 
$0.022$ \\
\hline\hline
 & & & $F_L$ & & \\
\hline
1.00--2.00  & 
$-$ & 
$-$ & 
$-$ &  $0.605_{-0.229 - 0.024}^{+0.179 + 0.021}$ 
  & $0.78$ \\
0.00--2.00      & 
$0.29^{+0.21}_{-0.18}\pm$0.02~ & 
$0.00~\,^{+\,0.13}_{-\,0.00}\,\pm 0.02$ & 
$0.30^{+0.16}_{-0.16}\pm 0.02$ &  
$0.323_{-0.178 - 0.020}^{+0.198 + 0.019}$ & 
$0.54$ \\
2.00--4.30      & 
$0.71^{+0.24}_{-0.24}\pm$0.05~ & 
$0.77\pm 0.15 \pm 0.03$ & 
$0.37^{+0.25}_{-0.24}\pm 0.10$ & 
 $0.754_{-0.198 - 0.018}^{+0.128 + 0.015}$  & 
$0.79$ \\
4.30--8.68    & 
$0.64^{+0.23}_{-0.24}\pm$0.07~ & 
$0.60~\,^{+\,0.06}_{-\,0.07}\,\pm 0.01$ & 
$0.68^{+0.15}_{-0.17}\pm 0.09$ & 
$0.634_{-0.216 - 0.022}^{+0.175 + 0.022}$ & 
$0.60$\\
10.09--12.89 & $0.17^{+0.17}_{-0.15}\pm$0.03~ & 
$0.41\pm 0.11 \pm 0.03$ & $0.47^{+0.14}_{-0.14}\pm 0.03$ & 
$0.482_{-0.208 - 0.013}^{+0.163 + 0.014}$ &   
$0.42$ \\
14.18--16.00    & 
$-0.15^{+0.27}_{-0.23}\pm$0.07~ & 
$0.37\pm 0.09\pm 0.05$ & 
$0.29^{+0.14}_{-0.13}\pm 0.05$ & 
$0.396_{-0.241 - 0.004}^{+0.141 + 0.004}$ & 
$0.36$ \\
$>$16.00       & 
0.12$^{+0.15}_{-0.13}\pm$0.02~ & 
$0.26~\,^{+\,0.10}_{-\,0.08}\,\pm 0.03$ & 
$0.20^{+0.19}_{-0.17}\pm 0.05$  & 
$0.357_{-0.133 - 0.003}^{+0.074 + 0.003}$ &  
$0.34$ \\
1.00--6.00      & 
$0.67^{+0.23}_{-0.23}\pm$0.05~ & 
$0.55\pm 0.10\pm 0.03$ & 
$0.69^{+0.19}_{-0.21}\pm 0.08$ & 
$0.703_{-0.212 - 0.019}^{+0.149 + 0.017}$ & 
$0.75$ \\
\hline\hline
\end{tabular}
\end{center}
\end{table*} 
 
\begin{table*}[htpb]
  \caption{Binned clean observables for $B \to K^* \mu^+ \mu^-$ 
(our numbers (CQM) vs. reference \cite{Descotes-Genon:2013vna}).}
\label{tab:bin-2}
  \begin{center}
    \begin{tabular}{c|cc|cc}
\hline\hline
Bin (GeV$^2$) & \cite{Descotes-Genon:2013vna} & CQM & \cite{Descotes-Genon:2013vna} & CQM \\
\hline
&  \multicolumn{2}{c|}{ $\av{P_1}$ } &  
                 \multicolumn{2}{c}{ $\av{P_2}$ }\\
\hline
1--2   & 
$0.007_{-0.005 - 0.051}^{+0.008 + 0.054}$ & 
$-0.0115773$ & 
$0.399_{-0.023 - 0.008}^{+0.022 + 0.006}$ & 
$0.47$ \\
0.1--2   & 
$0.007_{-0.004 - 0.044}^{+0.007 + 0.043}$ & 
$0.0108792$ & $0.172_{-0.009 - 0.018}^{+0.009 + 0.018}$ & 
$0.22$ \\
2.00--4.30   & 
$-0.051_{-0.009 - 0.045}^{+0.010 + 0.045}$ & 
$-0.266563$ & $0.234_{-0.085 - 0.016}^{+0.058 + 0.015}$ & 
$ 0.019$ \\
4.30--8.68   & 
$-0.117_{-0.002 - 0.052}^{+0.002 + 0.056}$ & 
$-0.372456$ & $-0.407_{-0.037 - 0.006}^{+0.048 + 0.008}$ & 
$-0.37$ \\
10.09--12.89 & 
$-0.181_{-0.361 - 0.029}^{+0.278 + 0.032}$ & 
$-0.470412$ & $-0.481_{-0.005 - 0.002}^{+0.08 + 0.003}$ & 
$-0.41$ \\
14.18--16.00 & 
$-0.352_{-0.467 - 0.015}^{+0.696 + 0.014}$ & 
$-0.614669$ & 
$-0.449_{-0.041 - 0.004}^{+0.136 + 0.004}$ & 
$-0.38$ \\
16.00--19 & 
$-0.603_{-0.315 - 0.009}^{+0.589 + 0.009}$ & 
$-0.777736$ & $-0.374_{-0.126 - 0.004}^{+0.151 + 0.004}$ & 
$-0.30$ \\ 	      			      
1.00--6.00   & 
$-0.055_{-0.008 - 0.042}^{+0.009 + 0.040}$ & 
$-0.26338$ & $0.084_{-0.076 - 0.019}^{+0.057 + 0.019}$ & 
$-0.060$ \\
\hline\hline
&  \multicolumn{2}{c|}{ $\av{P_3}$ } &  
                 \multicolumn{2}{c}{ $\av{P'_4}$ }\\
\hline
 1--2   & 
$-0.003_{-0.002 - 0.024}^{+0.001 + 0.027}$ & 
$0.00435836$ & $-0.160_{-0.031 - 0.013}^{+0.040 + 0.013}$ & 
$0.14$ \\
0.1--2   & 
$-0.002_{-0.001 - 0.023}^{+0.001 + 0.02}$ & 
$0.00159832$ & 
$-0.342_{-0.019 - 0.017}^{+0.026 + 0.018}$ & 
$-0.15$ \\
2.00--4.30   & 
$-0.004_{-0.003 - 0.022}^{+0.001 + 0.022}$ & 
$0.00454996$ & 
$0.569_{-0.059 - 0.021}^{+0.070 + 0.020}$ & 
$0.89$ \\
4.30--8.68   & 
$-0.001_{-0.001 - 0.027}^{+0.000+ 0.027}$ & 
$0.00224737$ & $1.003_{-0.015 - 0.029}^{+0.014 + 0.024}$ & 
$1.13$ \\
10.09--12.89 & 
$0.003_{-0.001 - 0.015}^{+0.000+ 0.014}$ & 
$0.00151139$ & 
$1.082_{-0.144 - 0.017}^{+0.140 + 0.014}$ & 
$1.21$ \\
14.18--16.00 & 
$0.004_{-0.001 - 0.002}^{+0.000+ 0.002}$ & 
$0.00101528$ & 
$1.161_{-0.332 - 0.007}^{+0.190 + 0.007}$ & 
$1.27$ \\
16.00--19 & 
$0.003_{-0.001 - 0.001}^{+0.001 + 0.001}$ & 
$0.00068909$ & 
$1.263_{-0.248 - 0.004}^{+0.119 + 0.004}$ & 
$1.33$ \\ 	      			     
1.00--6.00   & 
$-0.003_{-0.002 - 0.022}^{+0.001 + 0.020}$ & 
$0.00355465$ & 
$0.555_{-0.055 - 0.019}^{+0.065 + 0.018}$ & 
$0.83$ \\
\hline\hline
&  \multicolumn{2}{c|}{ $\av{P'_5}$ } &  
                 \multicolumn{2}{c}{ $\av{P'_8}$ }\\
\hline
 1--2   & 
$0.387_{-0.063 - 0.015}^{+0.047 + 0.014}$ & 
$0.258474$ & $-$ & 
$-0.039$ \\
0.1--2   & 
$0.533_{-0.036 - 0.020}^{+0.028 + 0.017}$ & 
$0.495414$ & $-$ & 
$-0.033$ \\
2.00--4.30   & 
$-0.334_{-0.111 - 0.019}^{+0.095 + 0.02}$ & 
$-0.423802$ & 
$-$ & 
$-0.026$ \\
4.30--8.68   & 
$-0.872_{-0.029 - 0.029}^{+0.043 + 0.03}$ & 
$-0.704599$ & 
$-$ & 
$-0.011$ \\
10.09--12.89 & 
$-0.893_{-0.110 - 0.017}^{+0.223 + 0.018}$ & 
$-0.697185$ & 
$-$ & 
$-0.0060$ \\
14.18--16.00 & 
$-0.779_{-0.363 - 0.009}^{+0.328 + 0.010}$ & 
$-0.600105$ & 
$-$ & 
$-0.0029$ \\
16.00--19 & 
$-0.601_{-0.367 - 0.007}^{+0.282 + 0.008}$ & 
$-0.449369$ & 
$-$ & 
$-0.0015$ \\ 	      			     
1.00--6.00   & 
$-0.349_{-0.098 - 0.017}^{+0.086 + 0.019}$ & 
$-0.394563$ & 
$-$ & 
$-0.023$ \\
\hline\hline
\end{tabular}
\end{center}
\end{table*}


\clearpage

\section{Summary}
\label{sec:summary}

We have performed a detailed analysis of the decay
process $B\to K^\ast(\to K\pi)\bar\ell\ell$ in the framework
of the covariant quark model by using the helicity formalism
to analyze the angular decay distribution.
All physical observables have been expressed in terms
of the helicity structure functions with taking into account
the effects of the finite lepton masses.  
We have found explicit relations between our helicity formalism and the approach based on the transversality amplitudes which is widely
used by both experimentalists and theorists. 
We have reported our numerical results on the branching fractions,
forward-backward asymmetry, longitudinal polarization and a set
of the so-called ``clean'' observables $P_i$ which depend on the hadron
uncertaintities (form factors) in a minimal way. Finally, 
we have compared the obtained results with available experimental data
and the results from other theoretical approaches.

\begin{acknowledgements}
We would like to thank our collaborators Th.~Gutsche, J.G.~K\"orner,
V.E.~Lyubovitskij and  P.~Santorelli for useful remarks and discussions.

The work was also partly supported by Slovak Grant Agency for Sciences VEGA, 
grant No. 1/0158/13 (S.~Dubni\v{c}ka, A.Z.~Dubni\v{c}kov\'{a}, A.~Liptaj), 
by Slovak Research and Development Agency APVV, 
grant No. APVV-0463-12 (S.~Dubni\v{c}ka, A.Z.~Dubni\v{c}kov\'{a}, A.~Liptaj) 
and by Joint research project of Institute of Physics, SAS and 
Bogoliubov Laboratory of Theoretical Physics, JINR, No. 01-3-1114 
(S.~Dubni\v{c}ka, A.Z.~Dubni\v{c}kov\'{a}, M.A.~Ivanov and A.~Liptaj).
\end{acknowledgements}

\end{document}